\documentclass[11pt]{article}

\usepackage{latexsym,amsmath,epsfig} 

\DeclareFontFamily{OT1}{rsfs10}{} 
\DeclareFontShape{OT1}{rsfs10}{m}{n}{ <-> rsfs10 }{} 
\DeclareMathAlphabet{\mathscript}{OT1}{rsfs10}{m}{n} 

\numberwithin{equation}{section}

\newcommand{\mbf}[1]{\mathbf{#1}}
\newcommand{\cohclass}[1]{\left[{#1}\right]}
\newcommand{\rest}[1]{\left.{#1}\right|}

\newcommand{\ns}{\normalsize}

\newcommand{\tr}{\text{tr}}

\newcommand{\w}{\wedge}
\newcommand{\Ds}{\not\!\!D}

\newcommand{\HdR}{H_{\text{DR}}}

\newcommand{\CC}{{\mathbf{C}}}
\newcommand{\ZZ}{{\mathbf{Z}}}
\newcommand{\RR}{{\mathbf{R}}}
\newcommand{\PP}{{\mathbf{P}}}

\def\d{\delta}

\def\l{\lambda}

\def\o{\omega}
\def\p{\pi}

\def\s{\sigma}

\def\cC{{\cal C}}
\def\cE{{\cal E}}
\def\cF{{\cal F}}

\def\cL{{\cal L}}

\def\cM{{\cal M}}
\def\cN{{\cal N}}
\def\cO{{\cal O}}

\begin{document}

%%%%%%%%%%%%%%%%%%%%%%%%%%%%%%%%%%%%%%%%%%%%%%%%%%%%%%%%%%%%%%%%%%%%%%

\begin{titlepage}

\vspace{-5cm}

\title{
   \hfill{\ns UPR-827T, OUTP-99-03P, PUPT-1834} \\[1em]
   {\LARGE Holomorphic Vector Bundles and Non-Perturbative 
    Vacua in M-Theory} \\[1em] } 
\author{
   Ron Donagi,$^1$
   Andr\'e Lukas$^2$, Burt A.~Ovrut$^3$
   and Daniel Waldram$^4$\\[0.5em]
   {\ns $^1$Department of Mathematics, University of Pennsylvania} \\[-0.4em]
   {\ns Philadelphia, PA 19104--6395, USA}\\ 
   {\ns $^2$Department of Physics, Theoretical Physics, 
       University of Oxford} \\[-0.4em]
   {\ns 1 Keble Road, Oxford OX1 3NP, United Kingdom}\\ 
   {\ns $^3$Department of Physics, University of Pennsylvania} \\[-0.4em]
   {\ns Philadelphia, PA 19104--6396, USA}\\ 
   {\ns $^4$Department of Physics, Joseph Henry Laboratories,}\\[-0.4em]
   {\ns Princeton University, Princeton, NJ 08544, USA}}
\date{}

\maketitle

\begin{abstract} 
We review the spectral cover formalism for constructing both $U(n)$
and $SU(n)$ holomorphic vector bundles on elliptically fibered
Calabi--Yau three-folds which admit a section. We discuss the allowed
bases of these three-folds and show that physical constraints
eliminate Enriques surfaces from consideration. Relevant properties of
the remaining del Pezzo and Hirzebruch surfaces are
presented. Restricting the structure group to $SU(n)$, we derive, in
detail, a set of rules for the construction of three-family particle
physics theories with phenomenologically relevant gauge groups. We
show that anomaly cancellation generically requires the existence of
non-perturbative vacua containing five-branes. We illustrate these
ideas by constructing four explicit three-family non-perturbative
vacua. 
\end{abstract}

\thispagestyle{empty}

\end{titlepage}

%%%%%%%%%%%%%%%%%%%%%%%%%%%%%%%%%%%%%%%%%%%%%%%%%%%%%%%%%%%%%%%%%%%%%%%

\section{Introduction}

The ground breaking work of Ho\v rava and Witten \cite{HW1,HW2} 
showed that $N=1$
supersymmetric, chiral theories can arise in four-dimensions upon
compactification of $M$-theory on an $S^{1}/Z_{2}$ orbifold times a
Calabi--Yau three-fold. Early work on this subject indicated that one could
get reasonable phenomenological values for Newton's constant and the gauge
unification parameter and scale \cite{W,bd}. 
Interestingly, acceptable values were
contingent upon the radius of the orbifold being an order of
magnitude, or more, larger than the Calabi--Yau radius. Thus, with decreasing
energy, the universe appears first eleven-, then five- and, finally,
four-dimensional. The effective four-dimensional reduction of Ho\v rava-Witten
theory was first constructed, directly from eleven-dimensions, in
\cite{low1,hp}. Various
aspects of this four-dimensional theory have been discussed by many authors
\cite{aq1}--\cite{bkl}.

More recently, the effective five-dimensional heterotic $M$-theory was
constructed \cite{losw1,losw2}. It was shown to be a 
specific form of gauged $N=1$
supergravity coupled to hyper and vector supermultiplets and bounded by two
four-dimensional orbifold fixed planes. These boundary planes contain $N=1$
supersymmetric gauge theories coupled to chiral matter supermultiplets. 
In \cite{losw1,losw2},
the so-called standard embedding of the spin connection into an $SU(3)$
subgroup of $E_{8}$ on one of the orbifold planes was assumed. It was shown in
\cite{losw1,losw2}, that this five-dimensional theory does 
not admit flat space as its static
vacuum solution. Rather, it supports BPS three-brane domain walls, much as
the gauged Type IIA supergravity theory discussed by Romans~\cite{lr} 
supports BPS eight-branes~\cite{8brane}. The minimal 
number of such domain walls is two, with one wall
located at each orbifold fixed plane. When expanded to leading non-trivial
order, this pair of BPS three-branes exactly reproduces the eleven-dimensional
``deformations'' of the metric line element discussed by Witten 
\cite{W}. We refer
to this as the minimal, or perturbative, vacuum. This five-dimensional theory
is not simply a formal development, since the universe actually passes through
this five-dimensional phase for an energy range  
of an order of magnitude or so below
the unification scale. Various physical aspects of this theory have been 
discussed in \cite{elpp,low4}. It was shown in 
\cite{losw1,losw2} that, when dimensionally reduced onto the
worldvolume of the pair of three-branes, the five-dimensional theory exactly
reproduces the effective four-dimensional theory derived by other methods in
\cite{low1,hp}, as it must.

As emphasized in \cite{nse}, the restriction to the standard 
embedding of the spin
connection into the gauge connection when reducing heterotic $M$-theory to
five- and four-dimensions is very unnatural. This is so because, unlike the
case of the weakly coupled heterotic superstring, there is no choice of
embedding, standard or otherwise, that is to say no choice of gauge field 
background, that allows one to set the entire supergravity three-form to
zero. Hence, in heterotic $M$-theory, one should consider arbitrary
gauge field backgrounds which preserve $N=1$ supersymmetry. 
This was also discussed in \cite{lpt}. 
In addition, as was first noted in~\cite{W}, one can include $M5$-branes
in the background and still preserve supersymmetry, provided the 
branes are wrapped on holomorphic curves within the Calabi--Yau threefold. 
These general vacua, which in five dimensions 
involving extra BPS three-branes, the remnants of the five-branes, in 
addition to the two located at the
orbifold planes, were also analyzed in \cite{nse}. 
It was shown that the worldvolume theories of the extra
three-branes are $N=1$ supersymmetric gauge theories, whose gauge
groups depend on the genus and the position in their moduli space of
the holomorphic curves. We refer to five-dimensional vacua with extra
BPS three-branes as non-perturbative vacua. The main conclusion of
\cite{nse} was that, because of the condition of anomaly cancellation,
including background five-branes greatly relaxes the constraint on
allowed non-standard embeddings, and allows much more freedom in
constructing vacua. It should be noted that there is a long history to
constructing backgrounds with non-standard embeddings ($(0,2)$
models), such as, for example,~\cite{dg} and the three-family model
of~\cite{kachru}, or more recently of~\cite{PR}. The new development
is the inclusion of five-branes in the vacuum. 

The results of \cite{nse} indicated the importance of heterotic
$M$-theories with non-standard embeddings and non-perturbative vacua,
analyzing the general structure of such backgrounds. A specific
example of these vauca was given recently in \cite{don1}, where
explicit constructions were carried out within the context of holomorphic
vector bundles on the orbifold planes of heterotic $M$-theory compactified on
elliptically fibered Calabi--Yau three-folds which admit a section. 
The results of \cite{don1} rely
upon recent mathematical work by Friedman, Morgan and Witten \cite{FMW}, 
Donagi \cite{D} and
Bershadsky, Johansen, Pantev and Sadov \cite{BJPS} 
who show how to explicitly construct such vector bundles, and on results
of \cite{cur,ba} who computed the family generation index 
in this context. Most recently work has also appeared discussing the stablity 
of this index under deformations of the bundle~\cite{DI}.
Extending
these results, we were able to formulate rules for constructing three-family
particle physics theories with phenomenologically interesting gauge groups. As
expected, the appearance of gauge groups other than the $E_{6}$ group of the
standard embedding, as well as the three-family condition, necessitate the
existence of $M5$-branes and, hence, non-perturbative vacua. In 
\cite{don1}, we showed
how to compute the topological class of these five-branes and, given this
class, analyzed an example of the moduli space of the associated holomorphic
curves. Our results were summarized as a set of rules for constructing vacua. 
In addition, we
gave one concrete example of a three-family model with gauge group
$SU(5)$, along with its five-brane class and moduli space.

In this paper, we greatly enlarge the discussion of the results in
\cite{don1},
deriving in detail the rules presented there. In order to make this work more
accessible to physicists, as well as to lay the foundation for the necessary 
derivations and proofs, we present brief discussions of (1) elliptically
fibered Calabi--Yau three-folds, (2) spectral cover constructions of both
$U(n)$ and $SU(n)$ bundles, (3) Chern classes and (4) complex
surfaces, specifically del Pezzo, Hirzebruch and Enriques
surfaces. Using this background, we 
explicitly derive the rules for the construction of three-family models based
on semi-stable holomorphic vector bundles with structure group $SU(n)$.
Specifically, we construct the form of the five-brane class $[W]$, as well as
the constraints imposed on this class due to the three-family condition, the
restriction that the vector bundle have structure group $SU(n)$ and the
requirement that $[W]$ be an effective class. From these considerations, we
derive a set of rules that are presented in section 7. As discussed in this
paper, elliptically fibered Calabi--Yau three-folds that admit a section can
only have del Pezzo, Hirzebruch, Enriques and blown-up Hirzebruch surfaces as
a base. We show in section 8, however, that Enriques surfaces can never
lead to effective five-brane curves in vacua with three generations.
Therefore, the base $B$ of the elliptic fibration is restricted to be a 
del Pezzo, Hirzebruch or a blow-up of a Hirzebruch surface. In Appendix B, we
present the generators of all effective classes in $H_{2}(B,\ZZ)$, as
well as the first and second Chern classes $c_{1}(B)$ and $c_{2}(B)$, for
these allowed bases. Combining the rules in section 7 with the generators
and Chern classes given in Appendix B, we present a general algorithm for the
construction of non-perturbative vacua corresponding to three-family particle
physics theories with phenomenologically relevant gauge groups. We illustrate
this algorithm by constructing four such non-perturbative vacua, three
with del Pezzo surfaces as a base and one with a Hirzebruch
surface. We do not, in this 
paper, discuss the moduli spaces of the five-brane holomorphic curves. An
explicit example of such a moduli space was given in \cite{don1}. A discussion
of the general method for constructing five-brane curve moduli spaces will be
presented, in detail, elsewhere \cite{wod}. We also leave for later
discussions of some natural phenomenological questions,
in particular the breaking of the gauge group down to the standard
model, since the main focus of this paper is to present the 
tools for constructing non-perturbative vacua.  
The results given here are all within the context of heterotic $M$-theory.
However, our formalism will apply, with very minor modifications, to
compactifications of the weakly coupled heterotic superstring on elliptically
fibered Calabi--Yau three-folds with $NS5$-branes. Finally, we would like to
point out that the five-brane classes $[W]$ that appear naturally in
three-family models have a component in the base surface. That is, they are
not wrapped purely on the fiber and, therefore, are not dual to three-branes
in $F$-theory. The interesting question of what vacua they are dual to in
$F$-theory and superstrings will be discussed elsewhere. In the context of 
toric varieties this has been addressed in an interesting paper by 
Rajesh~\cite{raj}.

%%%%%%%%%%%%%%%%%%%%%%%%%%%%%%%%%%%%%%%%%%%%%%%%%%%%%%%%%%%%%%%%%%%%%%%%%%%

\section{Holomorphic Gauge Bundles, Five-Branes and Non-Perturbative Vacua} 

In this section, we will discuss the generic properties of heterotic
$M$-theory vacua appropriate for a reduction of the theory to
$N=1$ supersymmetric theories in both five and four dimensions. 
The $M$-theory vacuum is
given in eleven dimensions by specifying the metric $g_{IJ}$ and the
three-form $C_{IJK}$ with field strength $G_{IJKL}=24\partial_{[I}
C_{JKL]}$ of the supergravity multiplet. Following Ho\v rava 
and Witten~\cite{HW1,HW2} and
Witten~\cite{W}, the space-time structure, to lowest
order in the expansion parameter $\kappa^{2/3}$, will be taken
to be 
\begin{equation}
   M_{11}=M_{4} \times S^{1}/Z_{2} \times X
\label{eq:1}
\end{equation}
where $M_{4}$ is four-dimensional Minkowski space, $S^1/Z_2$ is a
one-dimensional orbifold and $X$ is a smooth Calabi--Yau
three-fold. The vacuum space-time structure becomes more complicated
at the next order in $\kappa^{2/3}$, but this metric ``deformation'',
which has been the subject of a number of papers~\cite{W,bd,low1}, can be
viewed as arising as the static vacuum of the five-dimensional
effective theory~\cite{losw1,losw2} and, hence, need not concern us here. 

The $Z_2$ orbifold projection necessitates the introduction, on each
of the two ten-dimensional orbifold fixed planes, of an $N=1$, $E_8$ 
Yang-Mills supermultiplet which is required for anomaly cancellation. 
In general, one can consider vacua with non-zero gauge fields excited 
within the Calabi--Yau space, on each plane. However, the supersymmetry 
transformations imply the fields must be a solution of the hermitian 
Yang--Mills equations for an $E_8$-valued connection in order to be 
compatible with four preserved supercharges in four dimensions. 
Donaldson~\cite{Don} and Uhlenbeck and Yau~\cite{UhYau} have shown 
that picking a solution of the hermitian Yang--Mills equations is 
equivalent to the topological problem of choosing a semi-stable, 
holomorphic bundle with the structure group being the complexification 
$E_{8\mathbf{C}}$ of $E_8$. It is this second formulation we will use 
in this paper and we will often refer to fixing the background gauge 
fields as simply choosing a gauge bundle. In the following, we will
denote both the real and complexified groups by $E_8$, letting context 
dictate which group is being referred to. (In general, we will denote 
any group $G$ and its complexification $G_{\mathbf{C}}$ simply as $G$). 
These semi-stable, holomorphic gauge bundles are, a priori, allowed to be 
arbitrary in all other respects. In particular, there is no requirement 
that the spin-connection of the Calabi--Yau three-fold be embedded into 
an $SU(3)$ subgroup of the gauge connection of one of the $E_8$ bundles,
the so-called standard embedding. This generalization to arbitrary
semi-stable holomorphic gauge bundles is what is referred to as
non-standard embedding. The terms standard and non-standard embedding
are historical and somewhat irrelevant in the context of $M$-theory,
where no choice of embedding can ever set the entire three-form $C_{IJK}$ 
to zero. For this reason, we will avoid those terms and simply refer to
arbitrary semi-stable holomorphic $E_8$ gauge bundles. Fixing the gauge 
bundle will in general completely break the $E_8$ gauge symmetry in the 
low-energy theory. However, it is clear, that in order to preserve 
a non-trivial low-energy gauge group, we can restrict the
transition functions to be elements of any subgroup $G$ of $E_8$, such
as $G=U(n)$, $SU(n)$ or $Sp(n)$. We will refer to the restricted bundle as a
semi-stable, holomorphic $G$ bundle, or simply as a $G$ bundle. It is clear
that the $G_1$ bundle on one orbifold plane and the $G_2$ bundle on
the other plane need not, generically, have the same subgroups $G_1$
and $G_2$ of $E_8$. We will denote the semi-stable holomorphic gauge bundle
on the $i$-th orbifold plane by $V_i$ and the associated structure group
by $G_i$. 

In addition, as discussed in~\cite{W} and~\cite{nse,don1}, we will allow for 
the presence of five-branes located at points throughout the orbifold
interval. The five-branes will preserve $N=1$ supersymmetry
provided they are wrapped on holomorphic two-cycles within $X$ and
otherwise span the flat Minkowski space $M_4$~\cite{W,bbs,bsv}. The 
inclusion of five-branes is essential for a complete discussion of 
$M$-theory vacua. The reason for this is that, given a Calabi--Yau three-fold
background, the presence of five-branes allows one to construct large
numbers of gauge bundles that would otherwise be disallowed~\cite{nse,don1}. 

The requirements of gauge and gravitational anomaly cancellation on
the two orbifold fixed planes, as well as anomaly cancellation on each
five-brane worldvolume, places a further very strong constraint on,
and relationship between, the space-time manifold, the gauge bundles
and the five-brane structure of the vacuum. Specifically, anomaly
cancellation necessitates the addition of magnetic sources to the
four-form field strength Bianchi identity. The modified Bianchi identity 
is given by 
\begin{multline}
   \left(dG\right)_{11\bar{I}\bar{J}\bar{K}\bar{L}}
       =  2\sqrt{2}\pi \left(\frac{\kappa}{4 \pi}\right)^{2/3} 
             \left[ 2J^{(0)}\d(x^{11}) + 
                \right. \\ \left.
                2J^{(N+1)}\d(x^{11}-\pi\rho) +
                \sum_{n=1}^{N}J^{(n)}(\d(x^{11}-x_{n})+\d(x^{11}+x_{n}))
                \right]_{\bar{I}\bar{J}\bar{K}\bar{L}}
\label{eq:2}
\end{multline}
The sources $J^{(0)}$ and $J^{(N+1)}$ on the orbifold planes are 
\begin{equation}
   J^{(0)}= - \frac{1}{16\pi^2} \left.\left( 
              \tr F^{(1)} \w F^{(1)} - \frac{1}{2} \tr R \w R 
              \right) \right|_{x^{11}=0}
\label{eq:3}
\end{equation}
and
\begin{equation}
   J^{(N+1)} = - \frac{1}{16 \pi^{2}} \left.\left( 
              \tr F^{(2)} \w F^{(2)} - \frac{1}{2} \tr R \w R 
              \right) \right|_{x^{11}=0}
\label{eq:4}
\end{equation}
respectively. The two-form $F^{(i)}$ is the field strength of a
connection on the gauge bundle $V_{i}$ of the $i$-th orbifold plane
and $R$ is the curvature two-form on the Calabi-Yau three-fold. 
By ``$\tr$'' for the gauge fields we mean $\frac{1}{30}$-th of 
the trace in the $\mathbf{248}$ representation of $E_{8}$, while for the 
curvature it is the trace in the fundamental representation of the tangent 
space $SO(10)$. We have also introduced $N$ additional sources $J^{(n)}$, 
where $n=1,\dots,N$. These arise from $N$ five-branes located at
$x^{11}=x_{1},\dots,x_{N}$ where $0 \leq x_{1} \leq \cdots \leq x_{N}
\leq \pi\rho$. Note that each five-brane at $x=x_{n}$ has to be paired
with a mirror five-brane at $x=-x_{n}$ with the same source since the
Bianchi identity must be even under the $Z_{2}$ orbifold symmetry. These 
sources are four-form delta functions localized on the fivebrane world-volume. 
As forms there are Poincar\'e dual to the six-dimensional cycles of the 
fivebrane world volumes. (This duality is summarized in Appendix A). In 
particular their normalization is such that there are in integer cohomology 
classes. 

Non-zero source terms on the right hand side of the Bianchi
identity~\eqref{eq:2}  preclude the simultaneous vanishing of all
components of the three-form $C_{IJK}$. The result of this is that, to
next order in 
the Ho\v rava--Witten expansion parameter $\kappa^{2/3}$, the
space-time of the supersymmetry preserving vacua gets ``deformed''
away from that given in expression~\eqref{eq:1}. As discussed above,
this deformation of the vacuum need not concern us here. In this
paper, we will focus on yet another aspect of the Bianchi
identity~\eqref{eq:2}, a topological condition that constrains the
cohomology of the vacuum. This constraint is found as follows. Consider
integrating the Bianchi identity~\eqref{eq:2} over any five-cycle
which spans the orbifold interval together with an arbitrary
four-cycle ${\cal{C}}_{4}$ in the Calabi-Yau three-fold. Since $dG$ is
exact, this integral must vanish. Physically, this is the statement
that there can be no net charge in a compact space, since there is
nowhere for the flux to ``escape''. Performing the integral over the
orbifold interval, we derive, using~\eqref{eq:2}, that
\begin{equation}
   \sum_{n=0}^{N+1} \int_{{\cal{C}}_{4}} J^{(n)} = 0
\label{eq:5}
\end{equation}
Hence, the total magnetic charge over ${\cal{C}}_{4}$ vanishes. Since
this is true for an arbitrary four-cycle ${\cal{C}}_{4}$ in the
Calabi-Yau three-fold, it follows that the sum of the sources must be
cohomologically trivial. That is
\begin{equation}
   \cohclass{ \sum_{n=0}^{N+1} J^{(n)} } = 0
\label{eq:6}
\end{equation}
(Throughout this papar we will use the notation $[\o]$ to refer to the 
cohomology class of $\o$, in this case a closed four-form.) 
The physical meaning of this expression becomes more transparent if we
rewrite it using equations~\eqref{eq:3} and~\eqref{eq:4}. Using these
expressions, equation~\eqref{eq:6} becomes
\begin{equation}
   - \frac{1}{16\pi^{2}} \cohclass{\tr F^{(1)} \w F^{(1)}}
   - \frac{1}{16\pi^{2}} \cohclass{\tr F^{(2)} \w F^{(2)}}
   + \frac{1}{16\pi^{2}} \cohclass{\tr R \w R}
   + \sum_{n=1}^{N} \cohclass{J^{(n)}} = 0
\label{eq:7}
\end{equation}
It is useful to recall that the second Chern class of an arbitrary $G$ 
bundle $V$, thought of as an $E_{8}$ sub-bundle, is defined to be
\begin{equation}
   c_{2}(V)= - \frac{1}{16\pi^{2}} \cohclass{\tr F \w F}
\label{eq:8}
\end{equation}
Similarly, the second Chern class
of the tangent bundle of the Calabi-Yau manifold $X$ is given by
\begin{equation}
   c_{2}(TX)= - \frac{1}{16\pi^{2}} \cohclass{\tr R \w R}
\label{eq:9}
\end{equation}
where as above the trace is taken in the vector representation of 
$SO(6)\supset SU(3)$. It follows that expression~\eqref{eq:7} can be
written as  
\begin{equation}
   c_{2}(V_{1}) + c_{2}(V_{2}) +[W]= c_{2}(TX)
\label{eq:10}
\end{equation}
where 
\begin{equation}
   [W] = \sum_{n=1}^{N} [J^{(n)}]
 \label{eq:11}
\end{equation}
is the four-form cohomology class associated with the
five-branes. This is a fundamental constraint imposed on the vacuum
structure. We will explore this cohomology condition in great detail
in this paper. Since the Chern classes are integer, this is a condition  
between integer classes (or rather the image of $H^4(X,\ZZ)$ in $\HdR^4(X,\RR)$
as derived). This means that integrating this constraint over an arbitrary
four-cycle ${\cal{C}}_{4}$ yields the integral expression 
\begin{equation}
   n_{1}({\cal{C}}_{4}) + n_{2}({\cal{C}}_{4}) + n_{5}({\cal{C}}_{4})
    = n_{R}({\cal{C}}_{4})
\label{eq:12}
\end{equation}
which states that the sum of the number of gauge instantons on the two
orbifold planes, plus the sum of the five-brane magnetic charges, must
equal the instanton number for the Calabi-Yau tangent bundle, a number
which is fixed once the Calabi-Yau three-fold is chosen. Note that the 
normalizatio is such that one unit of five-brane charge is equal to one 
unit of instanton charge. 

To summarize, we are considering vacuum states of $M$-theory with the
following structure. 
\begin{itemize}
\item Space-time is taken to have the form 
\begin{equation}
  M_{11}=M_{4} \times S^{1}/Z_{2} \times X 
  \label{eq:last1}
\end{equation}
where $X$ is a Calabi-Yau three-fold.
\item There is a semi-stable holomorphic gauge bundle $V_{i}$ with fiber
group $G_{i} \subseteq E_{8}$ over the Calabi-Yau three-fold on the
$i$-th orbifold fixed plane for $i=1,2$. The structure groups $G_{1}$ and
$G_{2}$ of the two bundles can be any subgroups of $E_{8}$ and need
not be the same.
\item We allow for the presence of five-branes in the vacuum, which
are wrapped on holomorphic two-cycles within $X$ and are parallel to the 
orbifold fixed planes.
\item The Calabi-Yau three-fold, the gauge bundles and the five-branes
are subject to the cohomological constraint on $X$ 
\begin{equation}
  c_{2}(V_{1})+c_{2}(V_{2})+[W]=c_{2}(TX) 
  \label{eq:final2}
\end{equation}
where $c_{2}(V_{i})$ and
$c_{2}(TX)$ are the second Chern classes of the gauge bundle $V_{i}$
and the tangent bundle $TX$ respectively and $[W]$ is the class
associated with the five-branes .
\end{itemize}
Vacua of this type will be referred to as non-perturbative heterotic
M-theory vacua.

The discussion given in this section is completely generic, in that it
applies to any Calabi-Yau three-fold and any gauge bundles that can be
constructed over it. However, realistic particle physics theories
require the explicit construction of these gauge bundles. Until now,
such constructions have been carried out for the restricted cases
of standard or non-standard embeddings without
five-branes. These restrictions make it very difficult to obtain
realistic particle physics theories, that is, theories with three
families, appropriate gauge groups and so on. It is the purpose of
this paper to resolve these difficulties by
explicitly constructing theories with non-perturbative heterotic
$M$-theory vacua, utilizing the freedom introduced by including
five-branes. 

Specifically, we will present a formalism for the construction of
semi-stable holomorphic gauge bundles with fiber groups $G_{1}$ and $G_{2}$
over the two orbifold fixed planes. In this paper, for specificity, we
will restrict the structure groups to be
\begin{equation}
 G_{i}= U(n_{i}) \quad \mbox{or} \quad SU(n_{i})
 \label{eq:18}
\end{equation}
for $i=1,2$. Other structure groups, such as $Sp(n)$ or exceptional
groups, will be discussed elsewhere. Our explicit bundle constructions
will be achieved over the restricted, but rich, set of elliptically
fibered Calabi-Yau three-folds which admit a section. 
Such three-folds have been extensively discussed within the context of 
duality between string theory and $M$- and $F$-theory. Independent of 
this use, however, elliptically fibered Calabi--Yau three-folds with 
a section are known to be the simplest class of Calabi--Yau spaces
on which one can explicitly construct bundles, compute Chern classes,
moduli spaces and so on~\cite{FMW,D,BJPS}. This makes them a compelling
choice for the construction of concrete particle physics
theories. Having constructed the bundles, one can explicitly calculate
the gauge bundle Chern classes $c_{2}(V_{i})$ for $i=1,2$, as well as
the tangent bundle Chern class $c_{2}(TX)$. Having done so, one can
then find the class $[W]$ of the five-branes using the
cohomology condition~\eqref{eq:10}. That is, in this paper we will
present a formalism in which the structure of non-perturbative
$M$-theory vacua can be calculated. 

As will be discussed in detail below, having constructed a
non-perturbative vacuum, we can compute the number of low energy
families and the Yang-Mills gauge group associated with that
vacuum. We will show that, because of the flexibility introduced by
the presence of five-branes, we will easily construct non-perturbative
vacua with three-families. Similarly, one easily finds
phenomenologically interesting gauge groups, such as $E_{6}$, $SU(5)$
and $SO(10)$, as the $E_{8}$ subgroups commutant with the $G$-bundle structure
groups, such as $SU(3)$, $SU(4)$ and $SU(5)$ respectively, on the observable
orbifold fixed plane. In addition, using the cohomology
constraint~\ref{eq:10}, one can explicitly determine the cohomology
class $[W]$ of the five-branes for a specific vacuum. Hence, one can
compute the holomorphic curve associated with the five-branes exactly
and determine all of its geometrical attributes. These include the
number of its irreducible components, which tells us the number of 
independent five-branes, and its genus, which will tell
us the minimal gauge group on the five-brane worldvolume when dimensionally
reduced on the holomorphic curve. Furthermore, we are, in general,
able to compute the entire moduli space of the holomorphic curve. This
can tell us about gauge group enhancement on the five-brane
worldvolume, for example. In \cite{don1}, we 
discussed the generic properties of examples of holomorphic 
curves associated with five-branes. We will present a more 
detailed discussion in \cite{wod}.

Finally, we want to point out that there are more moduli associated
with these non-perturbative vacua. These are (1) the moduli associated
with the gauge instantons on the two orbifold planes and (2) the
translation moduli of the five-branes in the orbifold dimension. 
Taken along with the five-brane holomorphic curve moduli, these form 
an enormously complicated, but physically rich, space of 
non-perturbative vacua. The structure of the full moduli space of 
non-perturbative heterotic $M$-theory vacua will be discussed elsewhere.

%%%%%%%%%%%%%%%%%%%%%%%%%%%%%%%%%%%%%%%%%%%%%%%%%%%%%%%%%%%%%%%%%%%%%%%%%%%%

\section{Elliptically Fibered Calabi--Yau Three-Folds} 

As discussed previously, we will consider
non-perturbative vacua where the Calabi--Yau three-fold is an elliptic 
fibration which admits a section. In this section, 
we give an introduction to these spaces, 
summarizing the properties we will need in order to compute explicitly
properties of the vacua. 

An elliptically fibered Calabi--Yau three-fold $X$ consists of a base $B$,
which is a complex two-surface, and an analytic map
\begin{equation}
 \pi:X \to B
 \label{eq:add1}
\end{equation}
with the property that for a generic point $b \in B$, the fiber
\begin{equation}
 E_{b} = \pi^{-1}(b)
 \label{eq:add2}
\end{equation}
is an elliptic curve. That is, $E_{b}$ is a Riemann surface of genus one. In
addition, we will require that there exist a global section, denoted $\sigma$,
defined to be an analytic map
\begin{equation}
   \sigma:B \to X
\label{eq:add3}
\end{equation}
that assigns to every point $b \in B$ the zero element $\sigma(b)= p 
\in E_{b}$ discussed below. The requirement that the elliptic fibration have a
section is crucial for duality to $F$-theory and to make contact with the
Chern class formulas in \cite{FMW}. However, this assumption does not seem
fundamentally essential and we will explore bundles without sections in future
work \cite{wod2}.
The Calabi--Yau three-fold must be a complex K\"ahler
manifold. This implies that the base is itself a complex manifold, while we
have already assumed that the fiber is a Riemann surface and so has a
complex structure. Furthermore, the fibration must be holomorphic, that
is, it must have holomorphic transition functions. Finally, 
the condition that the Calabi--Yau three-fold has vanishing first Chern class
puts a further constraint on the types of fibration allowed.

Let us start by briefly summarizing the properties of an elliptic
curve $E$. It is a genus one Riemann surface and so can be embedded in the
two-dimensional complex projective space $\PP^{2}$. A simple way to
do this is by using the homogeneous Weierstrass equation
\begin{equation}
   zy^2 = 4x^3 - g_2 x z^2 - g_3 z^3
\label{Weier}
\end{equation}
where $x$, $y$ and $z$ are complex homogeneous coordinates on
$\PP^{2}$. It follows that we identify $(\l x,\l y,\l z)$ 
with $(x,y,z)$ for any
non-zero complex number $\l$. The parameters $g_2$ and $g_3$ encode the
different complex structures one can put on the torus. Provided $z\neq
0$, we can rescale to affine coordinates where $z=1$. We then see,
viewed as a map from $x$ to $y$, that there are two branch cuts in the
$x$-plane, linking $x=\infty$ and the three roots of the cubic equation
$4x^3-g_2x-g_3=0$. When any two of these points coincide, the elliptic
curve becomes singular. This corresponds to one of the cycles in the
torus shrinking to zero. Such singular behaviour is characterized by
the discriminant 
\begin{equation}
 \Delta=g_2^3-27g_3^2
 \label{eq:add4}
\end{equation}
vanishing. Finally, we note that the complex structure provides a
natural notion of addition of points on the elliptic curve. The torus
can also be considered as the complex plane modulo a discrete group of
translations. Addition of points in the complex plane then induces a
natural notion of addition of points on the torus. Translated to the
Weierstrass equation, the identity element corresponds to the point
where $x/z$ and $y/z$ become infinite. Thus, in affine coordinates, the
element $p \in E$ is the point $x=y=\infty$. This can be scaled elsewhere in
non-affine coordinates, such as to $x=z=0$, $y=1$.  

The elliptic fibration is defined by giving the elliptic curve $E$ over
each point in the base $B$. If we assume the fibration has a global
section, and in this paper we do, then on each coordinate 
patch this requires giving the
parameters $g_2$ and $g_3$ in the Weierstrass equation as functions
on the base. Globally, $g_2$ and $g_3$ will be sections of appropriate line
bundles on $B$. In fact, specifying the type of an elliptic fibration over $B$
is equivalent to specifying a line bundle on $B$. Given the elliptic fibration
$\pi :X \to B$, we define ${\cal{L}}$ as the line bundle on $B$
whose fiber at $b \in B$ is the cotangent line $T_{p}(E_{b})$ to the
elliptic curve at the origin. That is, ${\cal{L}}$ is the conormal
bundle to the section $\sigma(B)$ in $X$. Conversely, given
${\cal{L}}$, we take $x$ and $y$ to scale as
sections of ${\cal{L}}^{2}$ and ${\cal{L}}^{3}$ respectively, which means that
$g_2$ and $g_3$ should be sections of ${\cal{L}}^{4}$ and ${\cal{L}}^{6}$. By 
${\cal{L}}^{i}$ we mean the tensor product of the line bundle ${\cal{L}}$ with
itself $i$ times. In conclusion, we see that the elliptic fibration is
characterized by a line bundle ${\cal{L}}$ over the base $B$ together with a
choice of sections $g_2$ and $g_3$ of ${\cal{L}}^{4}$ and ${\cal{L}}^{6}$.

Note that the set of points in the base over which the fibration becomes
singular is given by the vanishing of
the discriminant $\Delta=g_2^3-27g_3^2$. It follows from the above discussion
that $\Delta$ is a section of the line
bundle  $\cL^{12}$. The   zeros of $\Delta$ then naturally define a
divisor, which in this case is a complex curve, in the base. Since $\Delta$ is
a section of $\cL^{12}$, the cohomology class of the discriminant curve
is 12 times the cohomology class of the divisors defined by sections
of $\cL$.

Finally, we come to the important condition that on a Calabi--Yau three-fold
$X$ the first Chern class of the tangent bundle $T_X$ must vanish. The
canonical bundle $K_X$ is the line bundle constructed as the determinant of
the holomorphic cotangent bundle of $X$. The condition that
\begin{equation}
   c_{1}(T_X)=0
\label{eq:ron1}
\end{equation}
implies that $K_{X}={\cal{O}}$, where ${\cal{O}}$ is the trivial bundle. This,
in turn, puts a constraint on $\cL$. To see this, note that the adjunction
formula tells us that, since $B$ is a divisor of $X$, the canonical bundle
$K_{B}$ of $B$ is given by
\begin{equation}
   K_{B}=\rest{K_{X}}_B \otimes N_{B/X}
\label{eq:ron2}
\end{equation}
where $N_{B/X}$ is the normal bundle of $B$ in $X$ and by $\rest{K_X}_B$ we 
mean the restriction of the canonical bundle $K_X$ to the base $B$. From the 
above discussion, we know that
\begin{equation}
N^{-1}_{B/X}=\cL, \qquad \rest{K_{X}}_B=\cO
\label{eq:ron3}
\end{equation}
Inserting this into~\eqref{eq:ron2} tells us that
\begin{equation}
 \cL=K_B^{-1}
 \label{eq:add5}
\end{equation}
This condition means that $K_{B}^{-4}$ and $K_{B}^{-6}$ must have sections 
$g_2$ and $g_3$ respectively. Furthermore, the Calabi--Yau property imposes
restrictions on how the curves where these sections vanish are allowed to
intersect. It is possible to classify the surfaces on which 
$K_{B}^{-4}$ and $K_{B}^{-6}$ have such sections. These are found to be
\cite{MV} the del Pezzo, Hirzebruch and Enriques surfaces, as well as blow-ups
of Hirzebruch surfaces. In this paper we will discuss the first three
possibilities in detail.

As noted previously, in order to discuss the anomaly cancellation 
condition, we will need
the second Chern class of the holomorphic tangent bundle of
$X$. Friedman, Morgan and Witten~\cite{FMW} show that it can be
written in terms of the Chern classes of the holomorphic tangent
bundle of $B$ as
\begin{equation}
   c_2(TX) = c_2(B) + 11 c_1(B)^2 + 12 \s c_1(B)
\label{tangc2}
\end{equation}
where the wedge product is understood, $c_{1}(B)$ and $c_{2}(B)$ are the first
and second Chern classes of $B$ respectively 
and $\sigma$ is the two-form Poincare dual to the global section. 
We have used the fact that 
\begin{equation}
  c_{1}(\cL)=c_{1}(K_{B}^{-1})=c_{1}(B)
\label{ron4}
\end{equation}
in writing~\eqref{tangc2}.

%%%%%%%%%%%%%%%%%%%%%%%%%%%%%%%%%%%%%%%%%%%%%%%%%%%%%%%%%%%%%%%%%%%%%%%%%%%

\section{Spectral Cover Constructions}

In this section, we follow the construction of semi-stable holomorphic
bundles on elliptically fibered Calabi--Yau manifolds presented 
in~\cite{FMW,D,BJPS}. The idea is to understand the
bundle structure on a given elliptic fiber and then to patch these
bundles together over the base. The authors in~\cite{FMW,D,BJPS} discuss a
number of techniques for constructing bundles with different gauge
groups. Here we will restrict ourselves to $U(n)$ and $SU(n)$
sub-bundles of $E_8$. These are sufficient to give suitable
phenomenological gauge groups. This restriction allows us to consider only the
simplest of the different constructions, namely that via spectral
covers. In this section, we will summarize the spectral cover construction,
concentrating on the properties necessary for an explicit discussion of
non-perturbative vacua. We note that for structure groups $ G \neq U(n)$ or 
$SU(n)$, the construction of bundles is more complicated than the construction
of rank $n$ vector bundles presented here.

As we have already mentioned, the condition of supersymmetry requires
that the $E_8$ gauge bundles admit a field strength satisfying the
hermitian Yang--Mills equations. Donaldson, Uhlenbeck and
Yau~\cite{Don,UhYau} have shown that this is equivalent to the
topological requirement that the associated bundle be semi-stable, with
transition functions in the complexification of the gauge group. Since
we are considering $U(n)$ and $SU(n)$ sub-bundles, this means
$U(n)_{\CC}=GL(n,\CC)$ and $SU(n)_{\CC}=SL(n,\CC)$ respectively. 
The spectral cover construction is given in terms of this latter
formulation of the supersymmetry condition. Note that the distinction 
between semi-stable and stable bundles corresponds to whether the
hermitian Yang-Mills field strength is reducible or not. This refers
to whether, globally, it can be diagonalized into parts coming from different
subgroups of the full gauge group. More precisely, it refers to
whether or not the holonomy commutes with more that just the center of the
group. Usually, a generic solution of the hermitian Yang--Mills
equations corresponds to a stable bundle. However, on some spaces, for
instance on an elliptic curve, the generic case is semi-stable. 

\subsection*{$U(n)$ and $SU(n)$ Bundles Over An Elliptic Curve}

We begin by considering semi-stable bundles
on a single elliptic curve $E$. A theorem of Looijenga~\cite{Looij} states
that the moduli space of such bundles for any simply-connected group
of rank $r$ is an $r$-dimensional complex weighted projective
space. For the simply-connected group $SU(n)$, this moduli space is 
the projective space $\PP^{n-1}$. $U(n)$ is not
simply-connected. $U(n)$ bundles have a discrete integer invariant,
their degree or first Chern class, 
which we denote by $d$. Let $k$ be the greatest common divisor of $d$ and $n$.
It can be shown that the moduli space of a $U(n)$ bundle of degree $d$ over a
single elliptic curve $E$ is the $k$-th symmetric product of $E$, denoted by
$E^{[k]}$. In this paper, we will restrict our discussion to $U(n)$ bundles of
degree zero. For these bundles, the moduli space is $E^{[n]}$.

A holomorphic $U(n)_\CC=GL(n,\CC)$ bundle $V$ over an elliptic curve
$E$ is a rank $n$ complex vector bundle. 
As discussed earlier, we will denote $U(n)_\CC$ simply as 
$U(n)$, letting
context dictate which group is being referred to.
To define the bundle, we need
to specify the holonomy; that is, how the bundle twists as one moves around
in the elliptic curve. The holonomy is a map from the fundamental
group $\pi_1$ of the elliptic curve into the gauge group. Since the fundamental
group of the torus is Abelian, the holonomy must map into the maximal
torus of the gauge group. This means we can diagonalize all the transition
functions, so that $V$ becomes the direct sum of line bundles (one-dimensional 
complex vector bundles)
\begin{equation}
  V=\cN_1\oplus\dots\oplus\cN_n
 \label{eq:add6}
\end{equation}
Furthermore, the Weyl group permutes the diagonal elements, so
that $V$ only determines the $\cN_{i}$ only up to permutations.
To reduce from a $U(n)$ bundle to an $SU(n)$ bundle, one imposes the
additional condition that the determinant of the transition functions
be taken to be unity. This implies that the product, formed by simply
taking the product of the transition functions for each bundle, satisfies
\begin{equation}
 \cN_1\otimes\dots\otimes\cN_n=\cO
 \label{eq:add7}
\end{equation}
where $\cO$ is the trivial bundle on $E$.
 
The semi-stable condition implies that the line bundles $\cN_i$ are of 
the same degree, which can be taken to be zero. We
can understand this from the hermitian Yang--Mills equations. On a
Riemann surface, these equations imply that the field strength is actually
zero. Thus, the first Chern class of each of the bundles $\cN_i$ must
vanish, or equivalently $\cN_i$ must be of degree zero. On an elliptic
curve, this condition means that there is a unique point $Q_i$ on $E$
such that there is a meromorphic section of $\cN_i$ which is allowed
to have a pole at $Q_i$ and is zero at the origin $p$. We can write
this as  
\begin{equation}
 \cN_{i}=\cO(Q_i)\otimes\cO(p)^{-1}
 \label{eq:add8}
\end{equation}
Let us briefly clarify this notation. One can associate a line bundle
to any divisor in a complex manifold $X$. A divisor is a subspace
defined locally by the vanishing of a single holomorphic function or
any linear combination of such subspaces $D=\sum_i a_i V_i$ with
integer coefficients $a_i$, and so is a space of one complex dimension
lower than $X$. The associated 
line bundle has a section corresponding to a meromorphic function on
$X$ with poles of order $\leq a_i$ on the subspaces $V_i$. (If $a_i$
is negative then the function has a zero of order $\geq -a_i$ on
$V_i$.) By $\cO(D)$ we mean the line bundle associated with the
divisor $D$. In particular, on a Riemann surface, the divisors are
collections of points and for example $\cO(Q_i)$ means the line bundle
with a section corresponding to a meromorphic function which is
allowed to have a first order pole at the particular point $Q_i$.

If one further restricts the structure group to be $SU(n)$, then 
condition~\eqref{eq:add7} translates into the requirement that 
\begin{equation}
 \sum_{i=1}^{n} (Q_i-p) =0 
 \label{eq:add9}
\end{equation}
where one uses the natural addition of points on $E$ discussed above.

Thus, on a given elliptic curve, giving a semi-stable $U(n)$ bundle
is equivalent to giving an unordered (because of the Weyl symmetry)
$n$-tuple of points on the curve. An $SU(n)$ bundle has the further
restriction that $\sum_i (Q_i-p)=0$. For an $SU(n)$ bundle,
these points can be represented very explicitly as roots of an equation
in the Weierstrass coordinates describing the elliptic curve. In affine
coordinates, where $z=1$, we write
\begin{equation}
   s = a_0 + a_2 x + a_3 y + a_4 x^2 + a_5 x^2y + \dots + a_n x^{n/2}
\label{speceq}
\end{equation}
(If $n$ is odd the last term is $a_nx^{(n-3)/2}y$.) Solving the equation $s=0$,
together with the Weierstrass equation (hence the appearance of only
linear terms in $y$ in $s$), gives $n$ roots corresponding to the $n$ points
$Q_i$, where one can show that $\sum_i(Q_i-p)=0$ as required.  One notes
that the roots are determined by the coefficients $a_i$ only up to an
overall scale factor. Thus the moduli space of roots $Q_i$ is the
projective space $\PP^{n-1}$ as anticipated, with the coefficients
$a_i$ acting as homogeneous coordinates. 

In summary, semi-stable $U(n)$ bundles on an elliptic curve are
described by an unordered $n$-tuple of points $Q_i$ on the elliptic
curve. $SU(n)$ bundles have the additional condition that 
$\sum_i(Q_i-p)=0$. In the $SU(n)$ case, these points can be realized as
roots of the equation $s=0$ and give a moduli space of bundles which
is simply $\PP^{n-1}$, as mentioned above.

\subsection*{The Spectral Cover and the Line Bundle ${\cal{N}}$}

Given that a bundle on an elliptic curve is described by the
$n$-tuple $Q_i$, it seems reasonable that a bundle on an
elliptic fibration determines how the $n$ points vary as
one moves around the base $B$. The set of all the $n$ points over the 
base is called the spectral
cover $C$ and is an $n$-fold cover of $B$ 
with $\pi_{C}: C \to B$. The spectral cover alone does not contain
enough information to allow us to construct the bundle $V$. To do this, one
must specify an additional line bundle, denoted by ${\cal{N}}$, on the
spectral cover $C$. One obtains ${\cal{N}}$, given the vector bundle $V$, as
follows. Consider the elliptic fiber $E_{b}$ at any point $b \in B$. It
follows from the previous section that
\begin{equation}
   \rest{V,E_{b}}= {\cal{N}}_{1b} \oplus \dots \oplus {\cal{N}}_{nb}
\label{eq:night1}
\end{equation}
where ${\cal{N}}_{ib}$ for $i=1,..,n$ are line bundles on $E_{b}$.
In particular, we get a decomposition of the fiber $V_{\sigma(b)}$ of $V$
at $p=\sigma(b)$. Let $V|_{B}$ be the restriction of $V$ to the base $B$
embedded in $X$ via the section $\sigma$. We have just shown that the
$n$-dimensional fibers of $V|_{B}$ come equipped with a decomposition into a
sum of lines. As point $b$ moves around the base $B$, these $n$ lines move in
one to one correspondence with the $n$ points $Q_{i}$ above $b$. This data
specifies a unique line bundle \footnote{When C is singular,
${\cal{N}}$ may be  
more generally a rank-$1$ torsion free sheaf on $C$. 
For non-singular $C$ this is the same as a 
line bundle. }
${\cal{N}}$ on $C$ such that the direct image
$\pi_{C*} {\cal{N}}$ is $V|_{B}$ with its given decomposition. 
The direct image $\pi_{C*} {\cal{N}}$ is a vector bundle on $B$ whose fiber at
a generic point $b$, where the inverse image $\pi_{C}^{-1}(b)$
consists of the $n$ distinct points $Q_{i}$, is the direct sum of the
$n$ lines ${\cal{N}}|_{Q_{i}}$.

\subsection*{Construction of Bundles}

We are now in a a position to construct the rank $n$ vector bundle starting
with the spectral data \cite{FMW,D,BJPS}. The spectral data consists 
of the spectral cover $C \subset X$ together with the line bundle
${\cal{N}}$ on $C$. The spectral cover 
is a divisor (hypersurface) $C \subset X$ which is of degree $n$ over the 
base $B$; that is, the restriction $\pi_{C} : C \to B$ 
of the elliptic fibration is an 
$n$-sheeted branched cover. Equivalently, the cohomology class of $C$ in 
$H^2(X,\ZZ)$ must be of the form
\begin{equation}
 [C] = n \sigma + \eta 
 \label{eq:day1}
\end{equation}
where $\eta$ is a class in $H^2(B,\ZZ)$ and $\sigma$ is the 
section. This is equivalent to saying that the line bundle ${\cal{O}}_{X}(C)$
on $X$ determined by $C$, whose sections are meromorphic functions on $X$ with
simple poles along $C$, is given by
\begin{equation}
 {\cal{O}}_{X}(C)= {\cal{O}}_{X}(n\sigma) \otimes {\cal{M}}
 \label{eq:hope1}
\end{equation}
where ${\cal{M}}$ is some line bundle on $X$ whose restriction to each fiber
$E_{b}$ is of degree zero. Written in this formulation
\begin{equation}
 \eta = c_{1}({\cal{M}})
 \label{eq:hope2}
\end{equation}
The line bundle ${\cal{N}}$ is, at this point, completely arbitrary.

Given this data, one can construct a rank $n$ vector bundle $V$ on
$X$. It is easy  to describe the restriction $V|_{B}$ of $V$ to the
base $B$. It is  simply the direct image $V|_{B}= \pi_{C*} {\cal{N}}$.
It is also easy to describe the restriction of $V$ to a general elliptic 
fiber $E_b$. Let $C\cap E_{b} = \pi_{C}^{-1}(b) = Q_1 + \ldots +Q_n$ and 
$\sigma \cap E_{b} = p$. Then each 
$Q_i$ determines a line bundle ${\cal{N}}_i$ of degree zero on $E_{b}$
whose sections are the meromorphic functions on $E_{b}$ with first order poles
at $Q_{i}$ which vanish at $p$. The restriction $V|_{E_{b}}$ is then
the sum of the ${\cal{N}}_i$. Now  
the main point is that there is a unique vector bundle $V$ on $X$ with 
these specified restrictions to the base and the fibers.

To describe the entire vector bundle $V$, we use the 
Poincare bundle ${\cal P}$. This is a 
line bundle on the fiber product $X \times _B X'$. Here $X'$ is the ``dual 
fibration'' to $X$. In general, this is another elliptic fibration which is 
locally, but not globally, isomorphic to $X$. However, when $X$ has a 
section (which we assume), then $X$ and $X'$ are globally isomorphic, so we 
can identify them if we wish. (Actually, the spectral cover $C$ lives most 
naturally as a hypersurface in the dual $X'$, not in $X$. When we described it 
above as living in $X$, we were implicitly using the identification of $X$ and 
$X'$.) The fiber product $X \times _B X'$ is a 
space of complex dimension four. It is fibered over $B$, the fiber
over $b \in B$ being the ordinary product $E_b \times {E'}_b$ of the
two fibers. Now, the Poincare bundle ${\cal P}$ is determined by the
following two properties: $(1)$ its restriction ${\cal P}|_{E_b \times
x}$ to a fiber $E_b \times x$, for $x \in {E'}_b$, is the line bundle
on $E_b$ determined by $x$ while $(2)$ its restriction to $\sigma
\times _B X'$ is the trivial bundle. Explicitly, ${\cal P}$ can be
given by the bundle whose sections are meromorphic functions on $X
\times _B X'$ with first order poles on ${\cal{D}}$ and which vanish
on $\sigma \times_{B} X'$ and on $X \times_{B} \sigma'$. That is
\begin{equation}
{\cal P} = {\cal O}_{X \times _B X'}({\cal{D}}-\sigma \times_{B} X'
           -X \times_{B} \sigma')\otimes K_{B}
 \label{eq:day2}
\end{equation}
where ${\cal{D}}$ is the diagonal divisor representing the graph of the
isomorphism $X\to X'$.

Using this Poincare bundle, we can finally describe the entire vector
bundle $V$ in terms of the spectral data. It is given by
\begin{equation} 
 V = {\p_1}_*({\p_2}^*{\cal{N}} \otimes {\cal P})
\label{eq:day3}
\end{equation}
Here $\p_1$ and $\p_2$ are the two projections of the fiber product $X
\times _B C$ onto the two factors $X$ and $C$. The two properties of
the Poincare bundle guarantee that the restrictions of this $V$ to the
base and the fibers indeed agree with the intuitive versions of $V_B$
and $V|_{E_b}$ given above. 

In general, this procedure produces $U(n)$ bundles. In order to get $SU(n)$ 
bundles, two additional conditions must hold. First, the condition 
that the line bundle
${\cal{M}}$ in equation~\eqref{eq:hope1} has degree 
zero on each fiber $E_{b}$ must be
strengthened to require that the restriction of  ${\cal{M}}$ to $E_{b}$ is the
trivial bundle. Hence, ${\cal{M}}$ is the pullback to $X$ of a line bundle on
$B$ which, for simplicity, we also denote by ${\cal{M}}$.
This guarantees that the restrictions to the fibers $V|_{E_b}$ are
$SU(n)$ bundles. The second condition 
is that $V|_{B}$ must be an $SU(n)$ bundle as well. That is, the line 
bundle ${\cal{N}}$ on $C$ is such that the first Chern class $c_1$ 
of the resulting bundle $V$ vanishes. This
condition, and its ramifications,  will be discussed in the next section.

$U(n)$ vector bundles on the orbifold planes of heterotic $M$-theory 
are always sub-bundles of an $E_{8}$ vector bundle. As such, issues arise
concerning their stability or semi-stability which are important and
require considerable analysis. Furthermore, the associated Chern classes
require an extended analysis to compute. For these reasons, in this paper, we
will limit our discussion to $SU(n)$ bundles, which are easier to study, and
postpone the important discussion of $U(n)$ bundles for a future publication
\cite{wod2}.

\subsection*{Chern Classes and Restrictions on the Bundle}

As discussed above, the global condition that the bundle be $SU(n)$ is that
\begin{equation}
 c_1(V)=(1/2\pi)\tr F=0
 \label{eq:add10}
\end{equation}
This condition is clearly true since, for structure group $SU(n)$,  
the trace must vanish. A formula for
$c_1(V)$ can be extracted from the discussion in Friedman, Morgan and
Witten \cite{FMW}. One finds that
\begin{equation}
   c_1(V) = \pi_{C*} \left( c_1(\cN) + \frac{1}{2}c_1(C) 
              - \frac{1}{2}\pi^*c_1(B) \right)
\end{equation}
where $c_1(B)$ means the first Chern class of the tangent 
bundle of $B$ considered as a complex vector bundle, and
similarly for $C$, while $\pi_{C}$ is the projection from the spectral
cover onto $B$; that is, $\pi_{C}:C\to B$. The operators $\pi_{C}^*$ and
$\pi_{C*}$ are the pull-back and push-forward of cohomology classes between $B$
and $C$. The condition that $c_1(V)$ is zero then implies that
\begin{equation}
   c_1(\cN) =  - \frac{1}{2}c_1(C) + \frac{1}{2}\pi_{C}^*c_1(B) + \gamma
\label{cN}
\end{equation}
where $\gamma$ is some cohomology class satisfying the equation
\begin{equation}
 \pi_{C*}\gamma=0
 \label{eq:add11}
\end{equation}
The general solution for $\gamma$ constructed from cohomology classes is
\begin{equation}
   \gamma = \l\left( n\s - \pi_{C}^* \eta + n \pi_{C}^* c_1(B) \right)
\label{gmma}
\end{equation}
where $\l$ is a rational number and $\s$ is the global section of the elliptic
fibration. Appropriate values for $\l$ will emerge shortly. 
From~\eqref{eq:day1} we recall that $c_1(C)$ is given by
\begin{equation}
    c_{1}(C)= -n\sigma -\pi_{C}^{*}\eta
\label{eq:add12}
\end{equation}
Combining the equations~\eqref{cN}, \eqref{gmma} and~\eqref{eq:add12} yields
\begin{equation}
   c_1(\cN) = n\left( \frac{1}{2} + \l \right) \s 
                + \left( \frac{1}{2} - \l \right) \pi_{C}^* \eta
                + \left( \frac{1}{2} + n\l \right) \pi_{C}^* c_1(B)
\end{equation}
Essentially, this means that the bundle $\cN$ is completely determined
in terms of the elliptic fibration and $\cM$. It is important to note, 
however, that there is not
always a solution for $\cN$. The reason for this is that $c_1(\cN)$ must be
integer, a condition that puts a substantial constraint on the allowed
bundles. To see this, note that the section is a horizontal divisor, 
having unit intersection
number with the elliptic fiber. On the other hand, the quantities $\pi_{C}
^*c_1(B)$ and
$\pi_{C}^*\eta$ are vertical, corresponding to curves in the
base lifted to the fiber and so have zero intersection number with the
fiber. Therefore, we cannot choose $\eta$ to cancel $\s$ and, hence,
the coefficient of $\s$ must, by itself, be an integer . This implies
that a consistent bundle $\cN$ will exist if either 
\begin{equation}
 \text{$n$ is odd},  \quad \l = m+\frac{1}{2} 
 \label{eq:add13}
\end{equation}
or
\begin{equation} 
   \text{$n$ is even},  \quad \l = m, \quad \eta = c_1(B) \bmod 2
 \label{eq:add14}
\end{equation}
where $m$ is an integer. Here, the $\eta = c_1(B) \bmod 2$ condition
means that $\eta$ and $c_1(B)$ differ by an even element of
$H^2(B,\ZZ)$. Note that when $n$ is even, we cannot
choose $\eta$ arbitrarily. These conditions are only sufficient for 
the existence of a consistent line bundle ${\cal{N}}$. They are also 
sufficient for all the examples we consider in this paper, and are the
only classes of solutions which is easy to describe in general. However, 
other solutions do exist. We could, for example take $n=4$, 
$\lambda=\frac{1}{4}$ and $\eta=2c_{1}(B)\pmod 4$, or $n=5$, $\l=\frac{1}{10}$ 
and $\eta=0\pmod 5$.

Finally, we can give the explicit Chern classes for the $SU(n)$ vector
bundle $V$. Friedman, Morgan and Witten calculate $c_1(V)$ and
$c_2(V)$, while Curio and Andreas~\cite{cur,ba} have found
$c_3(V)$. The results are  
\begin{equation}
\begin{align}
   c_1(V) &= 0 
      \label{c1} \\
   c_2(V) &= \eta \s - \frac{1}{24} c_1(B)^2 \left(n^3 - n\right) 
              + \frac{1}{2} \left(\l^2 - \frac{1}{4}\right) n \eta 
                      \left(\eta - nc_1(B)\right) 
     \label{c2} \\
   c_3(V) &= 2 \l \s \eta \left( \eta - nc_1(B) \right) 
     \label{c3}
\end{align}
\end{equation}
where the wedge product is understood.

%%%%%%%%%%%%%%%%%%%%%%%%%%%%%%%%%%%%%%%%%%%%%%%%%%%%%%%%%%%%%%%%%%%%%%%%

\section{Summary of Elliptic Fibrations and Bundles}

The previous two sections are somewhat abstract. For the sake of clarity, we
will here summarize those results which are directly relevant to
constructing physically acceptable non-perturbative vacua. First consider the 
Calabi--Yau space. 
\begin{itemize}

\item  An elliptically fibered Calabi--Yau three-fold is composed of a two-fold
base $B$ and elliptic curves $E_{b}$ fibered over each point $b \in B$. In
this paper, we consider only those elliptic fibrations that admit a
global section $\sigma$.

\item  The elliptic fibration is characterized by a single line bundle $\cL$
over $B$. The vanishing of the first Chern class of the canonical bundle
$K_{X}$ of the Calabi--Yau three-fold $X$ implies that 
\begin{equation}
 \cL= K_{B}^{-1}
 \label{eq:add15}
\end{equation}
where $K_{B}$ is the canonical bundle of the base $B$.

\item From the previous condition, it follows that the base $B$ is
restricted to del Pezzo, Hirzebruch and Enriques surfaces, as well as 
blow-ups of Hirzebruch surfaces. 

\item  The second Chern class of the holomorphic tangent bundle of $X$
is given by
\begin{equation}
   c_2(TX) = c_2(B) + 11 c_1(B)^2 + 12 \s c_1(B)
  \label{eq:add16}
\end{equation}
where $c_1(B)$ and $c_2(B)$ are the first and second Chern classes of $B$.

\end{itemize}

Next we summarize the spectral cover construction of semi-stable
holomophic gauge bundles.
\begin{itemize}

\item  A general semi-stable $SU(n)$ gauge bundle $V$ is determined by two
line bundles, $\cM$ and $\cN$. The relevant quantities associated 
with $\cM$ and $\cN$ are their first Chern classes
\begin{equation}
 \eta= c_{1}(\cM)
 \label{eq:add17}
\end{equation}
and $c_{1}(\cN)$ respectively. The class $c_{1}(\cN)$, in addition to
depending on $n$, $\sigma$, $c_{1}(B)$ and $\eta$, also contains a rational
number $\lambda$.

\item  The condition that $c_{1}(\cN)$ be an integer leads to the
sufficient but not necessary constraints on $\eta$ and $\lambda$ given by
\begin{equation}
\begin{aligned}
   {} & \text{$n$ is odd},  \quad \l = m+\frac{1}{2} \\
   {} & \text{$n$ is even},  \quad \l = m, \quad \eta = c_1(B) \bmod 2
\end{aligned}
\label{eq:add19}
\end{equation}
where $m$ is an integer.

\item  The relevant Chern classes of an $SU(n)$ gauge bundle $V$ are given by
\begin{equation}
\begin{align}
   c_1(V) &= 0 \label{c1V} \\
   c_2(V) &= \eta \s - \frac{1}{24} c_1(B)^2 \left(n^3 - n\right) 
              + \frac{1}{2} \left(\l^2 - \frac{1}{4}\right) n \eta 
                      \left(\eta - nc_1(B)\right) \label{c2V} \\
   c_3(V) &= 2 \l \s \eta \left( \eta - nc_1(B) \right) \label{c3V}
\end{align}
\label{eq:add20}
\end{equation}

\end{itemize}
How can one use the this data to construct realistic particle physics
theories? One proceeds as follows.
\begin{itemize}
\item  Choose a base $B$ from one of the allowed bases; namely, a del Pezzo,
Hirzebruch or Enriques surface, or a blow-up of a Hirzebruch surface. The
associated Chern classes $c_{1}(B)$ and $c_{2}(B)$ can be computed for any of
these surfaces.
\end{itemize}
This allows one to construct the second Chern class of the 
Calabi-Yau tangent bundle and a part of the gauge bundle 
Chern classes.
\begin{itemize}
\item  Specify $\eta$ and $\lambda$ subject to the
constraints~\eqref{eq:add19}. 
\end{itemize}
Given appropriate $\eta$ and $\lambda$, one can completely determine the 
relevant gauge bundle Chern classes. In addition we must satisfy the
cohomology  condition~\eqref{eq:10}. This relates the Chern classes to
the curves on which the five-branes wrap and it is what we will turn
to next.

%%%%%%%%%%%%%%%%%%%%%%%%%%%%%%%%%%%%%%%%%%%%%%%%%%%%%%%%%%%%%%%%%%%%%%%%%

\section{Effective Curves and Five-Branes}

Let us now consider the five-branes. Having discussed the Calabi--Yau 
three-fold $X$ and the gauge bundle, the third ingredient for defining a 
non-perturbative vacuum, is to give a set of five-branes wrapped on
holomorphic curves within $X$. We recall that the cohomology
condition~\eqref{eq:10} fixes the cohomology class associated with the
five-branes so that  
\begin{equation}
   [W]=c_{2}(TX)-c_{2}(V_{1})-c_{2}(V_{2})
\label{eq:x4}
\end{equation}
In order to make physical sense, this class must be Poincar\'e dual to the 
homology class of a set of curves in the Calabi--Yau space. As discussed in 
Appendix A, this means that $[W]$ must be effective. In general, this will 
restrict the bundles $V_i$ one can choose. To understand the form of this
restriction we need to find the set of effective curves on $X$. 

Consider a complex manifold $X$ which is an elliptic fibration over a
base $B$. Let us suppose we have found an effective class in
$H_2(B,\ZZ)$. Then, it naturally also lies in an effective homology
class in $H_2(X,\ZZ)$ of the elliptic fibration. Note that the
fibration structure guarantees that if two curves are in different
classes in the base, then they are in different classes in the full
manifold $X$. This implies, among other things, that if one finds the
effective generating class of the Mori cone of $B$, these classes
remain distinct classes of $X$. In addition, there is at least one
other effective class that is not associated with the base. This is
the class $F$ of the fiber itself. There may also be other effective
classes, for example, those related to points where the fiber
degenerates. However, we will ignore these since they will not appear
in the homology classes of the five-branes, our main interest in this
paper.  

The algebraic classes that do appear naturally are quadratic
polynomials in classes of the line bundles. The line bundle classes
are represented by two-forms or equivalently, under Poincar\'e
duality, by divisors, which are surfaces in $X$. Classes such as
$c_2(V_i)$, given in~\eqref{c2V}, are quadratic in these line bundle
classes. In terms of homology they are represented by curves
corresponding to the intersection of two divisors. The only line
bundle classes on a general elliptically fibered Calabi--Yau
three-fold $X$ are the base $B$ and the divisors
$\pi^{-1}({\cal{C}})$, where ${\cal{C}}$ is a curve in $B$. Any
quadratic polynomial in these classes can be written as
\begin{equation}
   W=W_{B} + a_{f}F
\label{decomp}
\end{equation}
where $ W_{B}$ is an algebraic homology class in the base manifold $B$ embedded
in $X$ and $a_{f}$ is some integer. Under what conditions is $W$ an 
effective class? It is clear that $W$ is effective if $W_{B}$ is an 
effective class in the base and $a_f\geq0$. One can also prove that
the converse is true in almost all cases. One sees this as
follows. First, unless a curve is purely in the fiber, in which case
$W_B=0$, the fact that $X$ is elliptically fibered means that all
curves $W$ project to curves in the base. The class $[W]$ similarly
projects to the class $W_B$. The projection of an effective class must
be effective, thus if $[W]$ is effective in $X$ then so is $W_B$ in
$B$. The only question then is whether there are effective curves in
$X$ with negative $a_f$. To address this we use the fact that any
effective curve must have non-negative intersection with any effective
divisor in $X$ unless the curve is contained within the divisor. The
intersections of $[W]$ with the effective divisor classes mentioned
above are given by
\begin{equation}
\begin{aligned}
   \pi_*\left( \pi^{-1}(\cC) \cdot W \right) &= \cC \cdot W_B \\
   \pi_*\left( B \cdot W \right) &= K_B \cdot W_B + a_f
\end{aligned}
\label{inters}
\end{equation}
where the intersections on the right-hand side are for classes in the
base. The second expression is derived by adjunction, recalling,
from~\eqref{eq:add5}, that the normal bundle to $B$ is
$\cN_{B/X}=\cL=-K_B$. From the first intersection one simply deduces
again that if $[W]$ is effective then so is $W_B$. Suppose that $a_f$
is non-zero. Then $W$ cannot be contained within $B$ and so from the
second expression we have $a_f\geq -K_B\cdot W_B$. From Appendix B we
recall that for del Pezzo and Enriques surfaces, $-K_B$ is nef, so
that its intersection with any effective class $W_B$ is
non-negative. Thus we do have $a_f\geq 0$ for $[W]$ to be
effective. The exception is a Hirzebruch surface for $r\geq 3$. We
then have $-K_B\cdot \cE =2>0$ but $-K_B\cdot S=2-r<0$. 

In conclusion, we see that, first $[W]$ is effective if and only if
$W_B$ is an effective class in $B$ and $a_f\geq 0$ for any del Pezzo
or Enriques surface. Second, this is also true for a Hirzebruch
surface $F_{r}$, with the exception of when $W_{B}$ happens to contain
the negative section $S$ and $r\geq3$. In this paper, for simplicity,
we will consider only those cases for which the statement is
true. Thus, under this restriction, we have that 
\begin{equation}
   W \text{ is effective } \Longleftrightarrow
   W_B \text{ is effective in } B \text{ and } a_f\geq 0
\label{thrm}
\end{equation}
This reduces the question of finding the effective curves in $X$ to 
knowing the generating set of effective curves in the base $B$. For
the set of base surfaces $B$ we are considering, finding such
generators is always possible. 

For simplicity, in this paper we will allow for arbitrary semi-stable 
gauge bundles $V_{1}$, which we henceforth call $V$, on the first 
orbifold plane, but always take the gauge bundle $V_{2}$ to be trivial. 
Physically, this corresponds to allowing observable sector gauge groups 
to be subgroups, such as $SU(5)$, $SO(10)$ or $E_{6}$, of $E_{8}$ but 
leaving the hidden sector $E_{8}$ gauge group unbroken. We do this only 
for simplicity. Our formalism also allows an analysis of the general case 
where the hidden sector $E_{8}$ gauge group is broken by a non-trivial 
bundle $V_{2}$. With this restriction, equation~\eqref{eq:x4} simplifies to
\begin{equation}
   [W]=c_{2}(TX)-c_{2}(V)
\label{eq:x5}
\end{equation}
Inserting the expressions~\eqref{eq:add16} and~\eqref{c2} for the
second Chern classes, we find that
\begin{equation}
   [W]= W_{B} +a_{f}F
\label{eq:x6}
\end{equation}
where
\begin{equation}
    W_{B}=\sigma( 12c_{1}(B)-\eta)
\label{eq:x7}
\end{equation}
is the part of the class associated with the base $B$ and
\begin{equation}
   a_f = c_2(B) + \left(11 + \frac{n^3-n}{24}\right) c_1(B)^2 
      - \frac{1}{2}n\left(\lambda^{2}-\frac{1}{4}\right)
          \eta\left(\eta-nc_{1}(B)\right)
\label{eq:x8}
\end{equation}
is the part associated with the elliptic fiber. 

As we have already stated, to make physical sense, $[W]$ must be an effective 
class. This physical requirement then implies, using the theorem~\eqref{thrm}, 
that necessarily
\begin{equation}
   W_B \text{ is effective in } B, \qquad
   a_f \geq 0
\label{eq:x9}
\end{equation}
This puts a further constraint on the allowed bundles in
non-perturbative vacua. Note, however, that this condition is much
weaker that the corresponding constraint without five-branes. In that
case $W_B$ and $a_f$ must vanish. It is this additional freedom which
greatly facilitates the construction of suitable particle physics
vacua.

%%%%%%%%%%%%%%%%%%%%%%%%%%%%%%%%%%%%%%%%%%%%%%%%%%%%%%%%%%%%%%%%%%%%%%%%%%

\section{Number of Families and Model Building Rules}

The first obvious physical criterion for constructing realistic particle
physics models is that we should be able to find theories with a small 
number of families, preferably three. We will see that this is, in fact, 
easy to do via the bundle constructions on elliptically fibered 
Calabi--Yau three-folds that we are discussing. We start by deriving 
the three family criterion as discussed, for instance, in Green, Schwarz 
and Witten~\cite{gsw}. The form of this condition for elliptically fibered 
Calabi--Yau manifolds was first given by Curio~\cite{cur}. 

The number of families is related to the number of zero-modes of the
Dirac operator in the presence of the gauge bundle on the Calabi--Yau
three-fold, since we want to count the number
of massless fermions of different chiralities. The original gauginos
are in the adjoint representation of $E_8$. In this paper, we are considering 
only gauge bundles $V$ with $SU(n)$ fiber groups. To count the number of
families, we need to count the number of fields in the matter
representations of the low energy gauge group, that is, the subgroup of
$E_{8}$ commutant with $SU(n)$, and their complex conjugates
respectively. Explicitly, in this paper, we will be 
interested in the following breaking patterns
\begin{equation}
\begin{aligned}
   E_8 \supset SU(3) \times E_6 : & \quad
      \mbf{248} = (\mbf{8},\mbf{1}) \oplus (\mbf{1},\mbf{78}) \oplus
           (\mbf{3},\mbf{27}) \oplus (\mbf{\bar{3}},\mbf{\bar{27}}) \\
   E_8 \supset SU(4) \times SO(10) : & \quad
      \mbf{248} = (\mbf{15},\mbf{1}) \oplus (\mbf{1},\mbf{45}) \oplus
           (\mbf{4},\mbf{16}) \oplus (\mbf{\bar{4}},\mbf{\bar{16}}) \oplus
           (\mbf{6},\mbf{10}) \\
   E_8 \supset SU(5) \times SU(5) : & \quad
      \mbf{248} = (\mbf{24},\mbf{1}) \oplus (\mbf{1},\mbf{24}) \oplus
           (\mbf{10},\mbf{5}) \oplus (\mbf{\bar{10}},\mbf{\bar{5}}) \oplus
           (\mbf{5},\mbf{\bar{10}}) \oplus (\mbf{\bar{5}},\mbf{10})
\end{aligned}
\label{subgroup} 
\end{equation}
Note, however, that the methods presented here will apply to any breaking
pattern with an $SU(n)$ subgroup.
We see that all the matter representations appear in the fundamental 
representation of the structure group $SU(n)$. By definition, 
the index of the Dirac operator
measures the difference in the number of positive and  negative
chirality spinors, in this case, on the Calabi--Yau three-fold. Since
six-dimensional chirality is correlated with four-dimensional
chirality, the index gives the number of families. From the fact that
all the relevant fields are in the fundamental representation of
$SU(n)$, we have that the number of generations is
\begin{equation}
   N_{\text{gen}} = \text{index}\,(V,\Ds) 
      = \int_X \text{td}\,(X) \text{ch}\,(V)
      = \frac{1}{2}\int_X c_3(V)
\label{eq:x10}
\end{equation}
where $\text{td}\,(X)$ is the Todd class of $X$. For the case of
$SU(n)$ bundles on elliptically fibered Calabi--Yau three-folds, 
one can show, using equation~\eqref{c3} above, that the number of 
families becomes
\begin{equation}
   N_{\text{gen}} = \l \eta ( \eta - nc_1(B) )
\label{eq:x11} 
\end{equation}
where we have integrated over the fiber. Hence, to obtain three families the
bundle must be constrained so that
\begin{equation}
  3 =  \lambda \eta \left( \eta - nc_1(B) \right)
\label{eq:x13} 
\end{equation}
It is useful to express this condition in terms of the class $W_{B}$ given in
equation~\eqref{eq:x7} and integrated over the fiber. We find that
\begin{equation}
 3= \lambda \left( W_{B}^{2}- (24-n)W_{B}c_{1}(B)
      + 12(12-n)c_{1}(B)^{2} \right)
\label{eq:x14}
\end{equation}
Furthermore, inserting the three family constraint into~\eqref{eq:x8} gives
\begin{equation}
   a_f = c_2(B) + \left(11 + \frac{1}{24}(n^3-n)\right) c_1(B)^2 
        - \frac{3n}{2\l}\left(\l^2-\frac{1}{4}\right)
\label{eq:x15}
\end{equation}

We are now in a position to summarize all the rules and constraints that are
required to produce particle physics theories with three families. We have 
that the homology class associated with the five-branes is specifically of
the form 
\begin{equation}
   [W]= W_{B} + a_{f}F
\label{eq:x16}
\end{equation}
where
\begin{gather}
   W_{B}=\sigma( 12c_{1}(B)-\eta)
      \label{eq:x17} \\
   a_f = c_2(B) + \left(11 + \frac{1}{24}(n^3-n)\right) c_1(B)^2
        - \frac{3n}{2\l}\left(\l^2-\frac{1}{4}\right)
      \label{eq:x18}
\end{gather}
and $c_{1}(B)$ and $c_{2}(B)$ are the first and second Chern classes of $B$.

The constraints for constructing particle physics vauca are then 
\begin{itemize}

\item  Effective condition: The requirement that $[W]$ is the class of a set 
of physical five-branes constrains $[W]$ to be an effective. 
Therefore, we must guarantee that
\begin{equation}
  W_{B} \text{ is effective in }B, \quad
   a_{f}\geq0  \text{ integer } 
 \label{eq:x19}
\end{equation}

\item  Three-family condition: The requirement that the theory have three 
families imposes the further constraint that 
\begin{equation}
 3= \lambda \left( W_{B}^{2}- (24-n)W_{B}c_{1}(B)
     + 12(12-n)c_{1}(B)^{2} \right)
\label{eq:x20}
\end{equation}

\end{itemize}
To these conditions, we can add the remaining relevant constraint from
section~4. It is
\begin{itemize}

\item  Bundle condition: The condition that $c_{1}({\cal{N}})$ be an
integer leads to the constraints on $W_{B}$ and $\lambda$ given by
\begin{equation}
\begin{aligned}
   {} & n \text{ is odd}, \quad \l = m+\frac{1}{2} \\
   {} & n \text{ is even}, \quad \l = m, \quad 
          W_B = {c_1(B) \bmod 2}
\end{aligned}
\label{eq:a21}
\end{equation}
where $m$ is an integer. Recall that this condition is sufficient, but not
necessary.

\end{itemize}
Note that in this last condition, the class $\eta$, which appeared in 
constraint~\eqref{eq:add19}, has been replaced by $W_{B}$. That this
replacement is valid can be seen 
as follows. For $n$ odd, there is no constraint on $\eta$ and, hence, 
using~\eqref{eq:x17}, no constraint on $W_{B}$. When $n$ is even, it 
is sufficient for $\eta $ to satisfy
$\eta=c_{1}(B) \bmod 2$. Since $12c_{1}(B)$ is an even element of
$H^{2}(B,\ZZ)$, it follows that $W_{B} = c_1(B) \bmod 2$.

It is important to note that all quantities and constraints have now been
reduced to properties of the base two-fold $B$. Specifically, if we know 
$c_1(B)$, $c_2(B)$, as well as a set of generators of 
effective classes in $B$ in which to expand
$W_{B}$, we will be able to exactly specify all appropriate non-perturbative
vacua. For the del Pezzo, Hirzebruch, Enriques and blown-up Hirzebruch
surfaces, all of these quantities are known.

Finally, from the expressions in~\eqref{subgroup} we find the following rule.
\begin{itemize}
 
\item  If we denote by $G$ the structure group of the gauge bundle and
by $H$ its commutant subgroup, then
\begin{displaymath}
 G=SU(3) \Longrightarrow  H=E_{6}
\end{displaymath}
\begin{equation}
 G=SU(4) \Longrightarrow  H=SO(10)
\label{eq:x24}
\end{equation}
\begin{displaymath}
 G=SU(5) \Longrightarrow  H=SU(5)
\end{displaymath}
$H$ corresponds to the low energy gauge group of the theory.

\end{itemize}
Armed with the above rules, we now turn to the explicit construction of
phenomenologically relevant non-perturbative vacua.

\section{Three Family Models}

In this section, we will construct four explicit solutions satisfying the
above rules.
In general, we will look for solutions where the class representing the
curve on which the fivebranes wrap is comparatively simple. As discussed
above, the allowed base surfaces $B$ of elliptically fibered Calabi--Yau
three-folds which admit a section are restricted to be the del Pezzo,
Hirzebruch and Enriques surfaces, as well as blow-ups of Hirzebruch surfaces.
Relevant properties of del Pezzo, Hirzebruch and Enriques surfaces, 
including their generators of effective curves, are given in the Appendix B. 
However, we now show that Calabi--Yau three-folds of this type  
with an Enriques base
never admit an effective five-brane curve if one requires that there be three
families. Recall that the cohomology class of the spectral cover must be of
the form
\begin{equation}
 [C]=n\sigma+ \eta
 \label{home1}
\end{equation}
and this necessarily is an
effective class in $X$. We may assume that $C$ does not contain $\sigma(B)$.
Otherwise, replace $C$ in the following discussion with its subcover $C'$
obtained by discarding the appropriate multiples of $\sigma(B)$.
This implies that the class of the intersection of $\s$ with $[C]$
\begin{equation}
 \sigma [C]=n\sigma^{2}+ \sigma\eta
 \label{home2}
\end{equation}
must be effective in the base $B$. Let us restrict $B$ to be an Enriques
surface. Using the adjunction formula, we find that
\begin{equation}
 \sigma^{2}=K_{B}
 \label{home3}
\end{equation}
where $K_{B}$ is the torsion class. Since $nK_{B}$ vanishes for even $n$, it
follows that when $n$ is even
\begin{equation}
 \sigma [C]= \sigma\eta
 \label{home4}
\end{equation}
Clearly, $\sigma\eta$ is effective, since $\sigma[C]$ is. For $n$ odd,
$nK_{B}=K_{B}$ and, hence
\begin{equation}
 \sigma [C]= K_{B}+ \sigma\eta
 \label{home5}
\end{equation}
Using the discussion in Appendix B, one can still conclude that $\sigma\eta$
is either an effective class or it equals $K_{B}$. From the fact that
\begin{equation}
 \sigma c_{1}(B)= K_{B}
 \label{home6}
\end{equation}
it follows, using equation~\eqref{eq:x17}, that the five-brane class
restricted to the Enriques base is given by
\begin{equation}
 W_{B}=12 K_{B}- \sigma\eta
 \label{home7}
\end{equation}
Since $12K_{B}$ vanishes, this becomes
\begin{equation}
 W_{B}=- \sigma\eta
 \label{home8}
\end{equation}
from which we can conclude that $W_{B}$ is never effective for non-vanishing
class $\sigma \eta$. Since, as explained above, $W_{B}$ must be effective for
the five-branes to be physical, such theories must be discarded. The only
possible loop-hole is when  $\sigma \eta$ vanishes or equals $K_{B}$. 
However, in this case, it
follows from~\eqref{eq:x11} that
\begin{equation}
 N_{\text{gen}} =0
 \label{home9}
\end{equation}
which is also physically unacceptable. We conclude that, on general grounds,
Calabi--Yau three-folds with an Enriques base never admit effective
five-brane curves if one requires that there be three families
\footnote{We thank E. Witten for pointing out to us the likelihood of this
conclusion.}. For this
reason, we henceforth restrict our discussion to the remaining possibilities.
In this section, for specificity, the base
$B$ will always be chosen to be either a del Pezzo surface or a Hirzebruch
surface. 

We first give two $SU(5)$ examples, each on del Pezzo surfaces; one
where the base component, $W_{B}$, is simple and one where the fiber component
has a small coefficient.

\subsection*{Example 1: $B=dP_{8}$, $H=SU(5)$}

We begin by choosing 
\begin{equation}
H=SU(5)
\label{eq:x25}
\end{equation} 
as the gauge group for our model. Then it follows from~\eqref{eq:x24}
that we must choose the structure group of the gauge bundle to be 
\begin{equation}
G=SU(5)
\label{eq:x26}
\end{equation}
and, hence, $n=5$. 

At this point, it is necessary to explicitly choose the base surface, which we
take to be
\begin{equation}
B=dP_{8}
\label{eq:x28}
\end{equation}
It follows from Appendix B that for the del Pezzo surface $dP_{8}$, a basis
for $H_{2}(dP_{8},\ZZ)$ composed entirely of effective classes 
is given by $l$ and $E_{i}$ for $i=1,..,8$ where
\begin{equation}
l \cdot l=1  \qquad  l \cdot E_{i}=0  \qquad E_{i} \cdot E_{j}=-\delta_{ij}
\label{eq:x29}
\end{equation}
There are other effective classes in $dP_{8}$ not obtainable as a linear
combination of $l$ and $E_{i}$ with non-negative integer coefficients, but we
will not need them in this example.
To these we add the fiber class $F$. Furthermore
\begin{equation}
c_{1}(B)= 3l- \sum_{r=1}^{8} E_{i}
\label{eq:x31}
\end{equation}
and
\begin{equation}
c_{2}(B)= 11
\label{eq:x32}
\end{equation}

We now must specify the component of the five-brane class in the base
and the coefficient $\l$ subject to the three
constraints~\eqref{eq:x19}, \eqref{eq:x20} and~\eqref{eq:a21}. Since
$n$ is odd, the bundle constraint~\eqref{eq:a21} tells us that
$\lambda=m+\frac{1}{2}$ for integer m. Here we will choose $m=1$ and
$W_B$ such that 
\begin{equation}
\begin{aligned}
   W_{B} &= 2E_{1}+E_{2}+E_{3} \\
  \lambda &=\frac{3}{2}
\label{eq:x33}
\end{aligned}
\end{equation}
Since $E_{1}$, $E_{2}$ and $E_{3}$ are effective, it follows that $W_{B}$ is
also effective, as it must be. Using the above intersection rules, one can
easily show that
\begin{equation}
W_{B}^{2}=-6,  \qquad  W_{B}c_{1}(B)=4,  \qquad
c_{1}(B)^{2}=1
\label{eq:x34}
\end{equation}
Using these results, as well as $n=5$ and $\lambda=\frac{3}{2}$, one finds 
that
\begin{equation}
   a_f = c_2(B) + \left(11 + \frac{n^3-n}{24}\right) c_1(B)^2 
        - \frac{3n}{2\l}\left(\l^2-\frac{1}{4}\right) = 17
\label{eq:x36}
\end{equation}
Since this is a positive integer, we have satisfied the effectiveness 
condition~\eqref{eq:x19} and the full five-brane class $[W]$ is effective 
in the Calabi--Yau three-fold $X$. Finally, we find that
\begin{equation}
\lambda( W_{B}^{2}- (24-n)W_{B}c_{1}(B)+12(12-n)c_{1}(B)^{2})=3
\label{eq:x35}
\end{equation}
and, therefore, the three family condition~\eqref{eq:x20} is satisfied.

This completes
our construction of this explicit non-perturbative vacuum. It represents a
model of particle physics with three families and gauge group $H=SU(5)$, along
with  explicit five-branes wrapped on a holomorphic curve with homology 
class
\begin{equation}
[W]=2E_{1}+E_{2}+E_{3} + 17F
\label{eq:x37}
\end{equation}
The properties of the moduli space of the five-branes were discussed in
\cite{don1} and will be explored in more detail in a future 
publication \cite{wod}.

\subsection*{Example 2: $B=dP_{9}$, $H=SU(5)$}

As a second example, we again choose gauge group 
\begin{equation}
H=SU(5)
\label{eq:x38}
\end{equation} 
and, hence,  the structure group 
\begin{equation}
G=SU(5)
\label{eq:x39}
\end{equation}
Then $n=5$ and we can again choose
$m=1$ and, therefore
\begin{equation}
\lambda=\frac{3}{2}
\label{eq:x40}
\end{equation}
In this example, we will take as a base surface\begin{equation}
B=dP_{9}
\label{eq:x41}
\end{equation}
It follows from Appendix B that a basis
for $H_{2}(dP_{9},\ZZ)$ composed entirely of effective classes 
is given by $l$ and $E_{i}$ for $i=1,..,9$. In addition, there are other
effective classes in $dP_{9}$ not obtainable as linear combinations of $l$ and
$E_{i}$ with non-negative integer coefficients. One such effective class is
\begin{equation}
c_{1}(B)= 3l- \sum_{r=1}^{9} E_{i}
\label{eq:x42}
\end{equation}
Furthermore
\begin{equation}
c_{2}(B)= 12
\label{eq:x43}
\end{equation}
We now must specify the component of the five-brane class in the base. 
In this example, we choose
\begin{equation}
W_{B}= 6E_{1}+E_{2}+E_{3}+12\left(3l-\sum_{i=1}^{9}E_{i}\right)
\label{eq:x44}
\end{equation}
Since $E_{1}$, $E_{2}$, $E_{3}$ and $3l-\sum_{i=1}^{9}E_{i}$ are effective, 
it follows that $W_{B}$ is
also effective, as it must be. Using the above intersection rules, one can
easily show that
\begin{equation}
W_{B}^{2}=154,  \qquad  W_{B}c_{1}(B)=8,  \qquad
c_{1}(B)^{2}=0
\label{eq:x45}
\end{equation}
Using these results, as well as $n=5$ and $\lambda=\frac{3}{2}$ one can check
that
\begin{equation}
\lambda\left( W_{B}^{2}- (24-n)W_{B}c_{1}(B)
    +12(12-n)c_{1}(B)^{2}\right)=3
\label{eq:x46}
\end{equation}
and, therefore, the three family condition is satisfied. Finally, let us
compute the coefficient $a_{f}$ of $F$. Using the above information, we find
that
\begin{equation}
   a_f = c_2(B) + \left(11 + \frac{n^3-n}{24}\right) c_1(B)^2 
        - \frac{3n}{2\l}\left(\l^2-\frac{1}{4}\right) = 2
\label{eq:x47}
\end{equation}
Since this is a positive integer, it follows from the above discussion that
the full five-brane curve $[W]$ is effective in the 
Calabi--Yau three-fold, as it must be. This completes
our construction of this explicit non-perturbative vacuum. It represents a
model of particle physics with three families and gauge group $H=SU(5)$, along
with  explicit five-branes wrapped on a holomorphic curve specified by 
\begin{equation}
[W]=6E_{1}+E_{2}+E_{3}+12(3l-\sum_{i=1}^{9}E_{i}) + 2F
\label{eq:48}
\end{equation}
Still within the context of del Pezzo base manifolds, we now give a
third example, this time with gauge group $H=SO(10)$. 

\subsection*{Example 3: $B=dP_{8}$, $H=SO(10)$}

In this third example, we choose the gauge group to be
\begin{equation}
H=SO(10)
\label{eq:x49}
\end{equation} 
and, hence,  the structure group 
\begin{equation}
G=SU(4)
\label{eq:x50}
\end{equation}
Then $n=4$. Since $n$ is even, then from constraint~\eqref{eq:a21} we must
have $\lambda=m$ where $m$ is an integer and $W_{B} = c_1(B)\bmod 2$. Here we
will choose $m=-1$ so that
\begin{equation}
\lambda=-1
\label{eq:x51}
\end{equation}
We will return to the choice of $W_{B}$ momentarily.
In this example, we will take as a base surface
\begin{equation}
B=dP_{8}
\label{eq:x52}
\end{equation}
Some of the effective generators and the first and second Chern
classes of $dP_{8}$ were given in the previous example.
We now must specify the component of the five-brane class in the base. 
In this example, we choose
\begin{equation}
W_{B}= 2E_{1}+2E_{2}+(3l-\sum_{i=1}^{8}E_{i})
\label{eq:x53}
\end{equation}
Since $E_{1}$, $E_{2}$ and $3l-\sum_{i=1}^{8}E_{i}$ are effective, 
it follows that $W_{B}$ is
also effective, as it must be. Furthermore, since 
\begin{equation}
c_{1}(B)= 3l- \sum_{r=1}^{8} E_{i}
\label{eq:x54}
\end{equation}
it follows that 
\begin{equation}
W_{B}= c_{1}(B) \bmod 2  
\label{eq:x55}
\end{equation}
since $2E_{1}+2E_{2}$ is an even element of $H^{2}(dP_{9},\ZZ)$. 
Using the above intersection rules, one can
easily show that
\begin{equation}
W_{B}^{2}=1,  \qquad  W_{B}c_{1}(B)=5,  \qquad
c_{1}(B)^{2}=1
\label{eq:x56}
\end{equation}
Using these results, as well as $n=4$ and $\lambda=-1$, one can check that
\begin{equation}
\lambda \left( W_{B}^{2}- (24-n)W_{B}c_{1}(B)
    +12(12-n)c_{1}(B)^{2} \right) = 3
\label{eq:x57}
\end{equation}
and, therefore, the three family condition is satisfied. Finally, let us
compute the coefficient $a_{f}$ of $F$. Using the above information, we find
that
\begin{equation}
   a_f = c_2(B) + \left(11 + \frac{n^3-n}{24}\right) c_1(B)^2 
        - \frac{3n}{2\l}\left(\l^2-\frac{1}{4}\right) = 29
\label{eq:x58}
\end{equation}
Since this is a positive integer, it follows from the above discussion that
the full five-brane curve $[W]$ is effective, as it must be. This completes
our construction of this explicit non-perturbative vacuum. It represents a
model of particle physics with three families and gauge group $H=SO(10)$, along
with  explicit five-branes wrapped on a holomorphic curve specified by 
\begin{equation}
[W]=2E_{1}+2E_{2}+(3l-\sum_{i=1}^{8}E_{i}) + 29F
\label{eq:59}
\end{equation}

\subsection*{Example 4: $B=F_r$, $H=SU(5)$}

We now return to choosing 
\begin{equation}
H=SU(5)
\label{eq:x60}
\end{equation} 
as the gauge group for our model. Then it follows from~\eqref{eq:x24} that 
we must choose the structure group of the gauge bundle to be
\begin{equation}
G=SU(5)
\label{eq:x61}
\end{equation}
and, hence, $n=5$. Since $n$ is odd, constraint~\eqref{eq:a21} tells us that
$\lambda=m+\frac{1}{2}$ for integer m. Here we will take $m=0$ and, therefore
\begin{equation}
\lambda=\frac{1}{2}
\label{eq:x62}
\end{equation}
In this example, we will choose the base surface to be a general Hirzebruch 
surface
\begin{equation}
B=F_r
\label{eq:x63}
\end{equation}
where $r$ is any non-negative integer.
It follows from Appendix B that for the Hirzebruch surface $F_r$, a basis
for $H_{2}(F_{r},\ZZ)$ composed entirely of effective classes 
is given by $S$ and ${\cal{E}}$ where
\begin{equation}
   {\cal{E}} \cdot {\cal{E}}=0,  \qquad 
   S \cdot S=-r,  \qquad  
   S \cdot {\cal{E}} =1  
\label{eq:x64}
\end{equation}
Furthermore, these completely generate the set of all effective classes.
To these classes we add the fiber class $F$. In addition
\begin{equation}
c_{1}(B) = 2S + (r+2){\cal{E}}
\label{eq:x65}
\end{equation}
and
\begin{equation}
c_{2}(B)= 4
\label{eq:x66}
\end{equation}
We now must specify the component of the five-brane class in the base. 
In this example, we choose
\begin{equation}
W_{B}= 26 S + (13r+23){\cal{E}}
\label{eq:x67}
\end{equation}
Since $S$ and ${\cal{E}}$ are effective, it follows that $W_{B}$ is
also effective, as it must be. Using the above intersection rules, one can
easily show that
\begin{equation}
W_{B}^{2}=1196,  \qquad  W_{B}c_{1}(B)=98,  \qquad
c_{1}(B)^{2}=8
\label{eq:x68}
\end{equation}
Note that the integer $r$ has cancelled out of these expressions.
Using these results, as well as $n=5$ and $\lambda=\frac{1}{2}$, one can check
that
\begin{equation}
\lambda( W_{B}^{2}- (24-n)W_{B}c_{1}(B)+12(12-n)c_{1}(B)^{2})=3
\label{eq:x69}
\end{equation}
and, therefore, the three family condition is satisfied. Finally, let us
compute the coefficient $a_{f}$ of $F$. Using the above information, we find
that
\begin{equation}
   a_f = c_2(B) + \left(11 + \frac{n^3-n}{24}\right) c_1(B)^2 
        - \frac{3n}{2\l}\left(\l^2-\frac{1}{4}\right) = 132
\label{eq:x70}
\end{equation}
Since this is a positive integer, it follows from the above discussion that
the full five-brane curve $[W]$ is effective, as it must be. This completes
our construction of this explicit non-perturbative vacuum. It represents a
model of particle physics with three families and gauge group $H=SU(5)$, along
with explicit five-branes wrapped on a holomorphic curve specified by 
\begin{equation}
[W] = 26S + (13r +23){\cal{E}} + 132F
\label{eq:x71}
\end{equation}
To repeat, we will explore the properties of the moduli spaces of
five-branes in detail in \cite{wod}.

%%%%%%%%%%%%%%%%%%%%%%%%%%%%%%%%%%%%%%%%%%%%%%%%%%%%%%%%%%%%%%%%%%%%%%%%%%%%

\appendix

\section*{Appendices}

\section{General Concepts}

\subsection*{Poincare Duality and Intersection Numbers}

The relationship between the cohomology class $[W]$ of the five-branes and
the holomorphic curves in $X$ over which they are wrapped arises from
the generic relationship between the cohomology and  homology groups
of a manifold. While this connection is a familiar one, since it is used 
extensively in this paper, we will give a brief description of it
here. 

Let $H_{k}(X,\RR)$ be the $k$-th real homology group over an
oriented manifold $X$. The elements of $H_{k}(X,\RR)$ are closed cycles
of dimension $k$ on
$X$.  Now, every element ${\cal{C}}_{k}$ of $H_{k}(X,\RR)$ can
be considered as a linear functional on forms in the de Rham
cohomology group $\HdR^{k}(X,\RR)$ in the following way. Let
$\phi$ be any element of $\HdR^{k}(X,\RR)$. Then  
\begin{equation}
 \phi \to \int_{{\cal{C}}_{k}} \phi
 \label{eq:13}
\end{equation}
defines a linear map from $\HdR^{k}(X,\RR)\to\RR$ 
for any ${\cal{C}}_{k}$ in $H_{k}(X,\RR)$. This map is well defined 
since two cycles ${\cal{C}}_{k}$ give the same integral if they are equal 
in homology (since $\phi$ is closed). In fact, all such linear maps can 
be realized this way and so the homology group $H_{k}(X,\RR)$ is dual 
as a vector space to $\HdR^{k}(X,\RR)$. This is simply the statement 
that homology and cohomology are dual to each
other. 

There is another notion of duality, Poincare duality, that
must be discussed to complete the story. Let $n$ be the real
dimension of the manifold $X$. One is then familiar with the notion 
of Poincar\'e duality for forms, that
the de Rham cohomology groups $\HdR^{k}(X,\RR)$ and
$\HdR^{n-k}(X,\RR)$ are dual as vector spaces, as follows. Let
$\phi$ be any element of $\HdR^{k}(X,\RR)$. Then 
\begin{equation}
 \phi \to \int_{X} \phi \wedge \psi
 \label{eq:14}
\end{equation}
defines a linear map from $\HdR^{k}(X,\RR)\to\RR$ for any 
$\psi$ in $\HdR^{n-k}(X,\RR)$. Again all such maps can be realized 
this way, and $\HdR^{k}(X,\RR)$ and $\HdR^{n-k}(X,\RR)$ are dual
vector spaces. Now, denote by $\eta_{{\cal{C}}_{k}}$ the element of
$\HdR^{n-k}(X,\RR)$ with the property that
\begin{equation}
 \int_{{\cal{C}}_{k}} \phi = \int_{X} \phi \wedge \eta_{{\cal{C}}_{k}} 
 \label{eq:15}
\end{equation}
for all $\phi \in \HdR^{k}(X,\RR)$. Then the mapping
\begin{equation}
 {\cal{C}}_{k} \to \eta_{{\cal{C}}_{k}}
 \label{eq:16}
\end{equation}
defines an isomorphism between the homology group $H_{k}(X,\RR)$ and 
the cohomology group $\HdR^{n-k}(X,\RR)$ since both are dual to 
$\HdR^k(X,\RR)$. This is the final result we want, that is 
\begin{equation}
 H_{k}(X,\RR) \cong  \HdR^{n-k}(X,\RR)
 \label{eq:17}
\end{equation}

For example, let $X$ be a Calabi-Yau three-fold and ${\cal{C}}_{2}$
any homology two-cycle contained in $H_{2}(X,\RR)$. Then, by
the above discussion, this two-dimensional cycle in $X$ can be
identified with a unique cohomology class $\eta_{{\cal{C}}_{2}}$
contained in $\HdR^{4}(X,\RR)$, and vice versa. This
expresses the exact relationship between the five-brane four-form $[W]$
and its associated holomorphic curve. 
We will refer to isomorphism~\eqref{eq:17} as the Poincare isomorphism and,
loosely speaking, to the pair ${\cal{C}}_{k}$ and $\eta_{{\cal{C}}_{k}}$ as
Poincare dual classes.
This isomorphism of forms and
homology classes is used extensively throughout this paper and we
often use the same notation for both objects in a Poincare dual pair.

Let $X$ be any oriented manifold of dimension $n$, $A$ an element of 
$H_{k}(X,\RR)$ and $B$ an element of $H_{n-k}(X,\RR)$. One can define the 
intersection number of $A$ and $B$ by taking representative cycles which 
intersect transversally, The intersection number is then the sum of 
intersections weighted with a plus or minus sign depending on the orientation 
of the intersection. In terms of the Poincar\'e dual forms it is given by
\begin{equation}
 A \cdot B = \int_{B} \eta_{A}= \int_{X} \eta_{A} \wedge \eta_{B}
 \label{eq:19}
\end{equation}
where $\eta_{A} \in \HdR^{n-k}(X,\RR)$ is the Poincare dual
of $A$ and $\eta_{B} \in \HdR^{k}(X,\RR)$ is the Poincare
dual of $B$. Since we often denote $\eta_{A}$ and
$\eta_{B}$ by $A$ and $B$ respectively, we can write
\begin{equation}
 A \cdot B =  \int_{X} A \wedge B
 \label{eq:20}
\end{equation}
Note that $A \cdot B = (-1)^{k(n-k)} B \cdot A$. A non-vanishing intersection
number $A \cdot B$ can be positive or negative, depending upon the
orientations of the tangent space basis vectors at the points of
intersection.

It is frequently essential in this paper to discuss the integer cohomology
groups $H_{k}(X,\ZZ)$.  There is a map from $H_{k}(X,\ZZ)\to H_{k}(X,\RR)$ 
whose kernel consists of torsion classes. If there is no torsion, the map 
is an embedding and all of the above statements are correct for
$H_{k}(X,\ZZ)$. If there is torsion, the above formulas still have
obvious analogues over $\ZZ$. 

\subsection* {Effective Curves and Homology}

Let $X$ be any $n$-dimensional complex manifold. A curve in $X$ is a closed 
subset which locally near each of its points can be defined by the
vanishing of $n-1$ (and no fewer) holomorphic functions. A curve is
irreducible if it is not the union of two proper subsets, each of which is
itself a curve. From now on we will take our manifold $X$ to be compact, 
that is, a
complex submanifold of a complex projective space. Then any curve in $X$ is
the union of a finite number of irreducible curves. To every curve
corresponds its homology class in $H_2(X,\ZZ)$. We say that a class
$\cC$ is irreducible if it is the class of an irreducible curve (though it 
may have other representatives which are reducible). We
say that a class $\cC$ is algebraic if it is a linear combination of
irreducible classes with integer coefficients.
That is, class ${\cal{C}}$ is
algebraic if
\begin{equation}
   {\cal{C}} = \sum_{i} a_{i}{\cal{C}}_{i}
\label{curvedef}
\end{equation}
where  ${\cal{C}}_{i}$ are irreducible classes and the coefficients 
$a_{i}$ are any integers. Note that when $X$ is a compact manifold the
sum is finite. The set of all algebraic classes, denote it by
$H_2(X,\ZZ)_{\text{alg}}$, forms a subgroup of $H_2(X,\ZZ)$. 

A class is called
effective if it is algebraic with all the coefficients
$a_{i}$ being non-negative.  One can show that there is
always a basis of  $H_2(X,\ZZ)_{\text{alg}}$ composed entirely 
of effective classes.  Clearly, any linear combination 
of such a basis with 
non-negative integer coefficients is also an effective class. Note, 
however, that there can be other effective classes not of this form. 
In general, the collection of all effective classes forms a cone in 
$H_2(X,\ZZ)_{\text{alg}}$ 
known as the Mori cone. The Mori cone can be shown to be linearly
generated by a set of  
effective classes. This set includes the effective basis of 
$H_2(X,\ZZ)_{\text{alg}}$
but is, in general, larger. The Mori cone can be finitely generated, as for
del Pezzo surfaces, or infinitely generated, as for $dP_{9}$ and 
Enriques surfaces. We refer
the reader to Appendix B for examples. 

By definition, any effective class
corresponds to a, in general reducible, curve in $X$. Non-effective
classes can not be interpreted as curves in $X$, since they involve
negative integers. Herein lies the importance of effective
classes. For example, in physical applications, such as the
five-branes in this paper, it is clearly essential that the classes
correspond to curves, as the five-branes must wrap around them. We,
therefore, must require five-brane classes to be effective. 

\section{Complex Surfaces}

\subsection*{Properties of del Pezzo Surfaces}

A del Pezzo surface is a complex manifold of complex dimension two
the canonical bundle of which is negative. This means that the dual 
anticanonical bundle has positive intersection with every curve in the 
surface. The del Pezzo surfaces which will concern us in this
paper are the surfaces $dP_r$
constructed from complex projective space $\PP^{2}$ by
blowing up $r$ points $p_{1},\dots,p_{r}$ in general position where
$r=0,1,\dots,8$.  

One also encounters the rational elliptic surface, which we denote $dP_9$, 
although it is not a del Pezzo surface in the strict sense. It can be obtained 
as the blow-up of $\PP^{2}$ at nine points which form the complete 
intersection of two cubic curves, and which are otherwise in general position.
For a $dP_9$ surface, the anticanonical bundle is no longer positive 
but, rather, it is ``nef'', 
which means that its intersection with every curve on the $dP_9$ surface is 
non-negative. In fact, a $dP_9$ surface is elliptically fibered 
over $\PP^1$ 
and the elliptic fibers (the proper transforms of the pencil of cubics through 
the nine blown up points) are in the anticanonical class. This
description fails when the nine points are in completely general
position, which is why we require them to be the complete intersection
of two (and, hence, of a pencil of) cubics. 

Of particular interest is the homology group of curves $H_{2}(dP_{r},\ZZ)$ 
on the del Pezzo surface. Since a new cycle is created each time a  point 
is blown up, we see that the dimension of
$H_{2}(dP_{r},\ZZ)$ is  $\dim H_2(dP_r,\ZZ)=r+1$. From $\PP^{2}$ we thus
inherit the single class of hyperplane divisors $l$. A representative of 
this class is any linear embedding of $\PP^{1}$ into $\PP^{2}$.  The
blow-up of the $i$-th point $p_{i}$  corresponds to an exceptional
divisor $E_{i}$. Hence, for $dP_{r}$, there are $r$ exceptional
divisors $E_{i}$, $i=1,\dots,r$. The curves $l$ and $E_{i}$ where
$i=1,\dots,r$ form a basis of homology classes of $H_{2}(dP_{r},\ZZ)$. 
Note that since $dP_{r}$ is a rational surface, $H^{2,0}(dP_{r})=0$ and, 
since on a surface, the Lefschetz theorem relates elements of $H^{1,1}$ to 
algebraic classes, we have $H_{2}(dP_{r},\ZZ)=H_{2}(dP_{r},\ZZ)_{\text{alg}}$. 
A particularly 
important element of $H_{2}(dP_{r},\ZZ)$ is the anticanonical class 
$\cF=-K_{dP_r}$, given by
\begin{equation}
 \cF=-K_{dP_r}=3l-\sum_{i=1}^{r}{E_{i}}.
 \label{eq:29}
\end{equation}

Let us consider the intersection numbers, defined in Appendix A, of the basis
of curves $l$ and $E_{i}$, $i=1,\dots,r$ of $H_{2}(dP_{r},\ZZ)$.
Now, any two lines in $\PP^2$ generically intersect once. Hence
one expects, and it can be shown, that
\begin{equation}
 l \cdot l= \int_{dP_{r}} l \wedge l= 1
 \label{eq:21}
\end{equation}
It is a known property of the exceptional divisors that each has self
intersection number $-1$. Furthermore, it is clear that exceptional
divisors associated with distinct points do not intersect. Therefore,
we have 
\begin{equation}
 E_{i} \cdot E_{j} = \int_{dP_{r}} E_{i} \wedge E_{j} = -\delta_{ij}
 \label{eq:22}
\end{equation}
Since a general line in $\PP^{2}$ does not pass through any of the 
blown up points, it follows that the proper transform of a
general line in $\PP^{2}$ does not intersect the $E_{i}$. Thus, we
have 
\begin{equation}
 E_{i} \cdot l= \int_{dP_{r}} E_{i} \wedge l= 0
 \label{eq:24}
\end{equation}

It is important to explicitly know the set of effective divisors on
$dP_{r}$. By definition, $l$ and $E_{i}$ for $i=1,\dots,r$ are
effective, as is the anticanonical class $\cF$. Now consider a line 
$l$ in $\PP^{2}$ which passes
through the $i$-th blown up point $p_{i}$. Such a line is still
effective. The class of such a curve is given by
\begin{equation}
 l-E_{i}
 \label{eq:25}
\end{equation}
and, hence, this is an effective divisor for any $i=1,\dots,r$. 
In general, a line can pass through at most two points, say $p_{i}$
and $p_{j}$ where $i\neq j$. The properties of blow-ups then imply
that the class of such a curve is 
\begin{equation}
 l-E_{i}-E_{j} 
 \label{eq:26}
\end{equation}
which, by construction, is an effective divisor for any $i\neq j =1,\dots,r$.

In general, a class $\cC \in H_{2}(dP_{r},\ZZ)$ is called exceptional if
it satisfies
\begin{equation}
\cC \cdot \cC=-1, \qquad \cC \cdot \cF=1
\label{eq:help1}
\end{equation}
where $\cF$ is the anticanonical class. 
The classes of the exceptional curves $E_i$ certainly are of this type, but 
there are others; for example, the class $l-E_{i}-E_{j}$ just described 
satisfies these properties as well. In fact, any exceptional class on
a general del Pezzo surface is the class of a unique, irreducible,
non-singular curve which can be blown down without creating any
singularities in the resulting surface. This curve is in fact a  
$\PP^1$ and has self-intersection $-1$. Such curves are called 
exceptional or simply $-1$ curves. Even though this is not apparent from our 
description, all these  $-1$ curves look exactly alike and, in fact, can be 
interchanged by the Weyl group which acts as a symmetry group of the family
of del Pezzo surfaces. So, for example, our del Pezzo surface 
admits another description, in which 
the line $l-E_{i}-E_{j}$ appears as the blow-up of some point, while
one or more of the exceptional divisors $E_{i}$ appears as a line or
higher degree curve. 

For $r \leq 4$, all exceptional curves are of the types already discussed. 
But consider, for $r \geq 5$, a conic in $\PP^{2}$; that is, a curve 
defined
by a quadratic equation. The conic is denoted by $2l$. A conic can
pass through at most five blown up points, say
$p_{i}$, $p_{j}$, $p_{k}$, $p_{l}$ and $p_{m}$. If they are all different, 
then the curves 
\begin{equation}
 2l-E_{i}-E_{j}-E_{k}-E_{l}-E_{m}
 \label{eq:27}
\end{equation}
are exceptional divisors. These are easily seen to be effective as
well. Similarly, consider a cubic in $\PP^{2}$; that is, a curve
defined by a cubic equation. The cubic is denoted by $3l$. When $r=7$,
$8$ or $9$, a cubic can be chosen to pass through one of the blown up
points, say $p_{i}$, twice (that is, it will be a singular cubic
curve, with singular point at $p_{i}$), while also passing (once)  
through six more of the blown up points, say 
$p_{j}$, $p_{k}$, $p_{l}$, $p_{m}$, $p_{n}$ and $p_{o}$. Therefore, we see 
that, for $r=7$, 8 or 9,
we also get exceptional divisors of the form 
\begin{equation}
 3l-2E_{i}-E_{j}-E_{k}-E_{l}-E_{m}-E_{n}-E_{o} 
 \label{eq:28}
\end{equation}
where all the points are different.  Again, these are easily seen to be
effective classes. Yet more examples of exceptional curves are 
obtained, for $r=8$ or $9$, by considering appropriate 
plane curves of degrees $4$, $5$ 
or $6$. The complete list of exceptional curves for $r \leq 8$ can be found, 
for example, in Table 3, page 35 of~\cite{math1}. All these classes are
effective.

We can now complete the description of the set of effective classes on a del 
Pezzo surface. These classes are precisely the linear combinations, 
with non-negative integer 
coefficients, of the anticanonical class $\cF$ and of the exceptional classes, 
including the $E_i$, the curves in~\eqref{eq:26}, \eqref{eq:27} and
\eqref{eq:28}, and their more complicated cousins for large $r$. For
$r \leq 8$ this gives us an explicit, finite set which generates the
Mori cone. 

The above statement, that is, that the effective classes are precisely the 
linear combinations, with non-negative integer coefficients, of the
anticanonical class $\cF$ and of the exceptional classes, remains true
for the rational elliptic surface $dP_9$. The new and, perhaps,
surprising feature is that on a $dP_9$ surface there are infinitely
many exceptional classes. This is easiest to see using the elliptic
fibration structure. Each of the nine exceptional divisors $E_i$ has
intersection number 1 with the elliptic fiber $\cF$, so it gives a
section of the fibration. Conversely, it is easy to see that any
section is an exceptional curve. But since each fiber, an elliptic
curve, is a group, it follows that the set of sections is itself a
group under the operation of pointwise addition of sections. We are
free to designate one of our nine sections, say $E_9$, as the ``zero''
section. The other eight sections then generate an infinite group of
sections, which generically will be $\ZZ^8$. The Mori cone in this
case is not generated by any finite set of effective curves.

Finally, we list the formulas for the Chern classes on $dP_{r}$. We find that 
\begin{equation}
 c_{1}(dP_{r})=-K_{dP_{r}}= 3l-\sum_{i=1}^{r}{E_{i}}
 \label{eq:29a}
\end{equation}
and
\begin{equation}
 c_{2}(dP_{r})=3+r
 \label{eq:30}
\end{equation}
are the first and second Chern classes of $dP_{r}$
respectively. The second Chern class is simply a number,  since there is
only one class in $H_{0}(dP_{r},\ZZ)$.

\subsection*{Properties of Hirzebruch Surfaces}

A Hirzebruch surface $F$ is a two-dimensional complex manifold constructed
as a fibration with base $\PP^1$ and fiber $\PP^1$. One way to construct these
surfaces is to start with a rank two vector bundle $V$ over $\PP^1$ and to
take $F$ to be the projectivization of $V$. For example, we can take $V$ to be
\begin{equation}
 V={\cal{O}}\oplus {\cal{O}}(r)
 \label{eq:111}
\end{equation}
where $r$ is a non-negative integer. 
The resulting Hirzebruch surface is denoted by $F_{r}$. It
is, in fact, easy to see that all Hirzebruch surfaces arise in this way. We
denote the fiber of $F_{r}$ over $\PP^1$ by ${\cal{E}}$. The sections
are not all equivalent. Let $S_{\infty}$ and $S_{0}$ denote the two
sections of $F_{r}$ corresponding to the sub-bundles ${\cal{O}}$ and
${\cal{O}}(r)$ respectively.The intersection numbers are found to be 
\begin{equation}
 {\cal{E}} \cdot {\cal{E}} =0, \qquad S_{\infty}\cdot S_{\infty}=-r, 
 \qquad S_{0}\cdot S_{0}= r
 \label{eq:222}
\end{equation}
and
\begin{equation}
 {\cal{E}} \cdot S_{\infty}= {\cal{E}} \cdot S_{0}= 1,  \qquad
 S_{\infty} \cdot S_{0}= 0
 \label{eq:333}
\end{equation}
These results are determined as follows.
Each section $S_{\infty}$ and $S_{0}$ meets each fiber ${\cal{E}}$ at
a unique point, while the self-intersection of a fiber is $0$ since it
can also be interpreted as the intersection of two distinct, hence
disjoint, fibers. The section $S_{\infty}$ corresponds to the lower
degree sub-bundle ${\cal{O}}$, so it can not be moved away from
itself. This is reflected in the negative self-intersection number. On
the other hand, $S_{0}$ corresponds to the larger bundle
${\cal{O}}(r)$. It moves in an $r$-dimensional linear system and any
two representatives meet in $r$ points. But a generic representative
of this system does not meet the section at infinity $S_{\infty}$,
thus providing the last intersection number. 

We should note, however, that some special curves in the linear system
will, in fact, meet $S_{\infty}$. These are forced to become reducible; that
is, they contain $S_{\infty}$ plus exactly $r$ fibers, leading to the equality
\begin{equation}
 S_{0}= S_{\infty}+ r{\cal{E}}
 \label{eq:444}
\end{equation}
which is valid in $H_{2}(F_{r},\ZZ)$. A basis for $H_{2}(F_{r},\ZZ)$ 
is provided by ${\cal{E}}$ together with either $S_{\infty}$ or $S_{0}$. 
The pair ${\cal{E}}$, $S_{\infty}$ has the advantage that it is also the 
set of generators for the Mori cone. That is, a class $a{\cal{E}}+bS_{\infty}$ 
is effective on $F_{r}$ for integers $a$ and $b$ if and only if
$a\geq0$ and $b\geq0$, as is easily seen from the intersection numbers 
above. 

In this paper, we will
choose ${\cal{E}}$ and $S_{\infty}$ as the basis of $H_{2}(F_{r},\ZZ)$.
Note again that since $F_{r}$ is a rational surface, $H^{2,0}(F_{r})=0$ and,
hence, $H_{2}(F_{r},\ZZ)=H_{2}(F_{r},\ZZ)_{\text{alg}}$.
If we denote $S_{\infty}$ simply by $S$, then the intersection numbers become
\begin{equation}
   {\cal{E}}  \cdot {\cal{E}} = 0, \qquad 
   S \cdot S = -r, \qquad 
   S\cdot {\cal{E}} = 1   
\label{eq:z1} 
\end{equation}
The first and second Chern classes are given by 
\begin{equation}
   c_1(F_r) = 2S + (r+2){\cal{E}}
 \label{eq:z2} 
\end{equation}
and
\begin{equation}
   c_2(F_r) = 4
\label{eq:z3}
\end{equation}
respectively.
Finally, to repeat, it is clear that 
$S$ and ${\cal{E}}$ are effective. Any other irreducible effective
curve must have a non-negative intersection number with $S$ and
${\cal{E}}$. From this condition, one finds that all effective curves
in $F_r$ are simply linear combinations of $S$ and ${\cal{E}}$ with
non-negative coefficients.

\subsection*{Properties of Enriques Surfaces}

Following \cite{math2}, we define an Enriques surface as a 
complex algebraic surface $B$ with $H^1(B,\CC)=0$, whose canonical bundle is 
torsion. That is
\begin{equation}
  K_B \neq {\cal O}_B, \qquad  K_B \otimes K_B = {\cal O}_B  
  \label{eq:555}
\end{equation}
It follows immediately from the definition that ${c_1(B)}^2=0$ and
$h^{2,0}=0$.  The Riemann-Roch theorem then implies that the 
Euler characteristic is $c_2(B) =12$, 
so $h^{1,1}=h^2=10$. In fact, the non-trivial cohomology is given by
\begin{equation}
  H^2(B,\ZZ) = \ZZ^{10} + \ZZ_2 
  \label{eq:xmas1}
\end{equation}
That is, the canonical bundle is the only torsion class. The 
intersection form on $H^2(B,\ZZ)$ vanishes on the torsion, while on the 
$\ZZ^{10}$ part 
it is even, unimodular and of signature (1,9). The torsion canonical bundle 
implies that the fundamental group of an Enriques surface is non-trivial and, 
in fact, is $\ZZ_2$. The universal cover is thus a double cover. It is a 
surface with Euler characteristic $2 \times 12 = 24$ and it has a trivial 
canonical bundle. It follows that the universal cover is a K3
surface. In other words, every Enriques surface is obtained as the
quotient of a K3 surface by an involution. This involution must act
freely, since the K3 is an unramified cover of the Enriques
surface. Since we require that the canonical bundle of the Enriques
surface is not the trivial bundle, it cannot have any global
sections. Thus, the involution on the K3 must send the holomorphic 
two-form to $-1$ times itself.

Although we will not use this in this paper, we mention the 
fact that the covering K3 is 
rather special. Among other properties, it must be elliptically fibered over 
$\PP^1$ and this fibration must also be preserved by the involution.
Therefore, the 
Enriques surface itself inherits a fibration by curves of genus $1$. However, 
we do not consider this to be an elliptic fibration, since it does not have a 
section. In fact, two of the fibers occur with multiplicity two, which
prevents the existence of a section even locally near these fibers. In
addition to these two double fibers, there are on a generic Enriques
surface exactly $12$ singular fibers, just as there are in the
elliptic fibration of a $dP_{9}$ surface. These two  
surfaces are actually related, the Enriques surface being obtained from a
$dP_{9}$ surface by 
performing logarithmic transforms along the two fibers which thereby become 
doubled. Conversely, the $dP_{9}$ surface can be recovered as 
the Jacobian fibration of the Enriques surface.

Since $H^{2,0}(B)=0$, all cohomology two-classes on the Enriques surface 
are algebraic. 
We need to decide which of these classes are effective. On a 
general Enriques surface it turns out that the effective classes fall into two 
components, each essentially the upper half of a ten-dimensional light cone. 
First we note that, by the adjunction formula, if $\cC$ is an
irreducible curve of arithmetic genus $g$ on the Enriques surface,
then the self-intersection number is
\begin{equation}
  \cC^2 =2g-2 \geq -2 
  \label{eq:xmas2}
\end{equation}
with equality holding if and only if $\cC$ is a smooth rational curve. 
Some special Enriques 
surfaces may certainly contain such smooth, rational curves, 
but not the general Enriques 
surface, as can be seen by a deformation argument. Therefore, 
we are left to discuss 
irreducible curves of non-negative self-intersection. Let us ignore
torsion for the moment and, hence, consider $H^2(B,\RR)$. The cone in
$H^2(B,\RR)$ of all classes of non-negative self-intersection looks
like the time-like cone of 10D Lorentzian geometry; that is, it 
consists of two components, the ``past'' and the ``future'' 
(recall that the signature is 
(1,9)). Any ample class $h$ takes positive values on one side of the cone
and negative values on 
the other. Therefore, all the effective classes are in one half of the cone. 
Conversely, we claim that all integral classes in this half cone are
effective.  This follows from the fact that for any class $\cC$ with
$\cC^2 \geq 0$, exactly one of the two classes $\cC$ or $-\cC$ is
effective, as can be seen from the Riemann-Roch formula. Since, in our
case, $-\cC$ lies in the wrong half cone it cannot be effective and,
therefore, $\cC$ must be effective.  

So far we ignored the torsion by considering $H^2(B,\RR)$. Returning to 
$H^2(B,\ZZ)=\ZZ^{10}+\ZZ_2$, we see that along 
with each class $\cC$ comes 
another class, $K_B+\cC$, with the same image in $H^2(B,\RR)$. 
Fortunately, with a 
single exception, these are both effective (or not) together, so the 
effectivity of a class $\cC \in H^2(B,\ZZ)$ depends only on its image in 
$H^2(B,\RR)$. The single exception is, of course, the pair $0$ and $K_B$ 
itself. (We are still assuming that our Enriques surface is general,
so we only consider classes satisfying $\cC^2 \geq 0$.) The reasoning
is similar to the above; that is, the Riemann-Roch theorem tells us
that either $\cC$ or $K_B-\cC$ must be effective and, likewise, that
either $K_B+\cC$ or $-\cC$ must be effective. But $\cC$ and $-\cC$
cannot both be effective (unless $\cC=0$), nor can $K_B-\cC$,
$K_B+\cC$ (unless $\cC=K_B$), so if $\cC$ is effective, so must be
$K_B+\cC$ and vice versa. 

For more information on Enriques surfaces we refer the reader to 
\cite{math2} or \cite{math3}.

\subsection*{Acknowledgements}

We would like to thank Ed Witten for helpful discussions. R.D. and B.A.O.
would like to thank A. Grassi and T. Pantev for useful conversations.
R.D. is supported in
part by an NSF grant DMS-9802456 as well as a University of Pennsylvania
Research Foundation Grant. 
A.L. is supported by the European Community under contract No. FMRXCT 960090.
B.A.O. is supported in part by a Senior Alexander von Humboldt Award, 
by the DOE under contract No. DE-AC02-76-ER-03071 and by a University 
of Pennsylvania Research Foundation Grant. 
D.W. is supported in part by the DOE under contract
No. DE-FG02-91ER40671.

%%%%%%%%%%%%%%%%%%%%%%%%%%%%%%%%%%%%%%%%%%%%%%%%%%%%%%%%%%%%%%%%%%%%%%

%%%%%%%%%%%%%%%%%%%%%%%%%%%%%%%%%%%%%%%%%%%%%%%%%%%%%%%%%%%%%%%%%%%%%%%%%%%%

\end{document}